\documentclass[acmtocl,acmnow]{acmtrans2m}

\newdef{definition}[theorem]{Definition}
\newdef{remark}[theorem]{Remark}

\usepackage{graphicx}
\usepackage{url}
\usepackage{amsmath}
\usepackage{algorithm}
\usepackage[noend]{algorithmic}

\graphicspath{{images//}} 

\title{Overlapping Community Detection in Networks: the State of the Art and Comparative Study\footnote{Citation: JIERUI XIE, STEPHEN KELLEY and BOLESLAW K. SZYMANSKI, Overlapping Community Detection in Networks: the State of the Art and Comparative Study, ACM Computing Surveys, vol. 45, no. 4, 2013 (In press)}}

\author{JIERUI XIE (jierui.xie@gmail.com)\\ Network Science and Technology, Rensselaer Polytechnic Institute, New York, USA\\
STEPHEN KELLEY (kelleys@ornl.gov) \\Oak Ridge National Laboratory, Tennessee, USA \and 
BOLESLAW K. SZYMANSKI (szymansk@cs.rpi.edu)\\ Network Science and Technology, Rensselaer Polytechnic Institute, New York, USA}

\begin{abstract} 
This paper reviews the state of the art in \textit{overlapping} community detection algorithms, quality measures, and benchmarks. A thorough comparison of different algorithms (a total of fourteen) is provided. In addition to \textit{community level} evaluation, we propose a framework for evaluating algorithms' ability to detect \textit{overlapping nodes}, which  helps to assess \textit{over-detection} and \textit{under-detection}. After considering community level detection performance measured by Normalized Mutual Information, the Omega index, and node level detection performance measured by F-score, we reached the following conclusions. For low overlapping density networks, SLPA, OSLOM, Game and COPRA offer better performance than the other tested algorithms. For networks with high overlapping density and high overlapping diversity, both SLPA and Game provide relatively stable performance. However, test results also suggest that the detection in such networks is still not yet fully resolved. A common feature observed by various algorithms in real-world networks is the relatively small fraction of overlapping nodes (typically less than 30\%), each of which belongs to only 2 or 3 communities.  

\end{abstract}

\category{A.1}{General Literature}{INTRODUCTORY AND SURVEY}%

\category{I.5.3}{Clustering}{Clustering}[Algorithms]

\category{H.3.3}{Clustering}{Information Search and Retrieval}[Clustering]
  
\category{E.1}{Data}{DATA STRUCTURES}[Graphs and networks]

\terms{Algorithms, Performance}     
       
\keywords{Algorithms, overlapping community detection, social networks}
\begin{document}
\maketitle


\section{Introduction}
\label{sec:intr}

Community or modular structure is considered to be a significant property of real-world social networks as it often accounts for the functionality of the system. Despite the ambiguity in the definition of \textit{community}, numerous techniques have been developed for both efficient and effective community detection. Random walks, spectral clustering, modularity maximization, differential equations, and statistical mechanics have all been used previously.   Much of the focus within community detection has been on identifying \textit{disjoint} communities. This type of detection assumes that the network can be partitioned into dense regions in which nodes have more connections to each other than to the rest of the network. Recent reviews on disjoint community detection are presented in \cite{Danon05comparingcommunity,LancichinettiComp:2009,wwwLeskovec:2010,Santo:2010}. 

However, it is well understood that people in a social network are naturally characterized by \textit{multiple} community \textit{memberships}. For example, a person usually has connections to several social groups like family, friends, and colleagues; a researcher may be active in several areas. Further, in online social networks, the number of communities an individual can belong to is essentially unlimited because a person can simultaneously associate with as many groups as he wishes. This also happens in other complex networks such as biological networks, where a node might have multiple functions. In \cite{goldbergchapter6:2011,Fergal:2011}, the authors showed that the \textit{overlap} is indeed a significant feature of many real-world social networks. 

For this reason,  there is growing interest in overlapping community detection algorithms that identify a set of clusters that are not necessarily disjoint. There could be nodes that belong to more than one cluster. In this paper, we offer a review on the state of the art in this area.

\section{Preliminaries}
In this section, we present basic definitions that will be used throughout the paper. Given a network or graph $G=\{E,V\}$, $V$ is a set of $n$ nodes and $E$ is a set of $m$ edges. For dense graphs $m=O(n^2)$, but for sparse networks $m=O(n)$. The network structure is determined by the $n\times n$ adjacency matrix $A$ for unweighted networks and weight matrix $W$ for weighted networks.  Each element $A_{ij}$ of $A$ is equal to 1 if there is an edge connecting nodes $i$ and $j$; and it is 0 otherwise. Each element $w_{ij}$ of W takes a nonnegative real value representing strength of connection between nodes $i$ and $j$.

In the case of overlapping community detection, the set of clusters found is called a \textit{cover}  $C=\{c_1, c_2, \cdots, c_k\}$ \cite{LancichinettiNMI-LFM:2009}, in which a node may belong to more than one cluster. Each node $i$ associates with a community according to a \textit{belonging factor} (i.e., soft assignment or membership) $\left[a_{i1}, a_{i2}, \cdots, a_{ik}\right]$  \cite{Nepusz:2008}, in which $a_{ic}$ is a measure of the strength of association between node $i$ and cluster $c$. Without loss of generality, the following constraints are assumed to   be satisfied
\begin{equation}
\label{eq:belonging}
		0\leq a_{ic}\leq 1 \hspace{0.2cm} \forall i \in V, \forall c \in C
\end{equation}
and 
$$\sum_{c=1}^{|C|} a_{ic}=1,$$ where $|C|$ is the number of clusters. However, the belonging factor is often solely a set of artificial weights. It may not have a clear or unambiguous physical meaning \cite{EAGLE:2009b}.

In general, algorithms produce results that are composed of one of two types of assignments, \textit{crisp}  (non-fuzzy) assignment or \textit{fuzzy} assignment \cite{SteveSurvey:2011}. With crisp assignment, the relationship between a node and a cluster is \textit{binary}. That is, a node $i$ either belongs to cluster $c$ or does not. With fuzzy assignment, each node is associated with communities in proportion to a belonging factor. With a threshold, a fuzzy assignment can be easily converted to a crisp assignment. Most detection algorithms output crisp community assignments.

\section{Algorithms}
\label{sec:rw}
In this section, algorithms for overlapping community detection are reviewed and categorized into five classes which reflect how communities are identified.

\subsection{Clique Percolation}
The clique percolation method (CPM) is based on the assumption that a community consists of overlapping sets of fully connected subgraphs and detects communities by searching for adjacent cliques.  It begins by identifying all cliques of size $k$ in a network.  Once these have been identified, a new graph is constructed such that each vertex represents one of these $k$-cliques.  Two nodes are connected if the $k$-cliques that represent them share $k-1$ members.  Connected components in the new graph identify which cliques compose the communities.  Since a vertex can be in multiple $k$-cliques simultaneously, overlap between communities is possible. CPM is suitable for networks with dense connected parts. Empirically, small values of $k$ (typically between $3$ and $6$) have been shown to give good results \cite{CPM:2005,LancichinettiComp:2009,COPRA:2010}. CFinder\footnote{\url{http://www.cfinder.org.}} is the implementation of CPM, whose time complexity is polynomial in many applications \cite{CPM:2005}. However, it also fails to terminate in many large social networks. 

CPMw \cite{CPMwFarkas:2007} introduces a subgraph intensity threshold for weighted networks. Only $k$-cliques with intensity larger than a fixed threshold are included into a community. Instead of processing all values of $k$, SCP \cite{SCPKumpula:2008} finds clique communities of a given size. In the first phase, SCP detects $k$-cliques by checking all the ($k-2$)-cliques in the common neighbors of two endpoints when links are inserted to the network sequentially in the order of decreasing weights. In the second phase, the $k$-community is detected by finding the connected components in the ($k-1$)-clique projection of the bipartite representation, in which one type of node represents a $k$-clique and the other denotes a ($k-1$)-clique. Since each $k$-clique is processed exactly twice, the running time grows linearly as a function of the number of cliques. SCP allows multiple weight thresholds in a single run and is faster than CPM.

Despite their conceptual simplicity, one may argue that CPM-like algorithms are more like pattern matching rather than finding communities since they aim to find specific, localized structure in a network.

\subsection{Line Graph and Link Partitioning}
The idea of partitioning links instead of nodes to discover community structure has also been explored. A node in the original graph is called overlapping if links connected to it are put in more than one cluster. 

In \cite{YYLinkClustering:2010}\footnote{\url{https://github.com/bagrow/linkcomm.}}, links are partitioned via hierarchical clustering of edge similarity.  Given a pair of links $e_{ik}$ and $e_{jk}$ incident on a node $k$, a similarity can be computed via the Jaccard Index defined as
 $$S(e_{ik},e_{jk})=\frac{|N_{i}\cap N_{j}|}{|N_{i}\cup N_{j}|},$$ where $N_{i}$ is the neighborhood of node $i$ including $i$. Single-linkage hierarchical clustering is then used to build a link dendrogram. Cutting this dendrogram at some threshold yields link communities. The time complexity is $O(nk_{max}^2)$, where $k_{max}$ is the maximum node degree in the network.

Evans \cite{LineGraph2:2009,LineGraph1:2010} projected the network into a weighted \textit{line graph}, whose nodes are the links of the original graph. Then disjoint community detection algorithms can be applied. The node partition of a line graph leads to an edge partition of the original graph. CDAEO \cite{Zhihao:2010} provides a post-processing procedure to determine the extent of overlapping. Once the preliminary partitioning on the line graph is done, for a node $i$ with $|E_{icmin}|/|E_{icmax}|$ below some predefined threshold, where $E_{icmin(cmax)}$ is the set of edges in the community with which $i$ has the minimum (maximum) number of connections, links in $E_{icmin}$ of the line graph are removed. This essentially reduces node  $i$ to a single membership. 

Kim \cite{Youngdokim:2011} extended the map equation method (also known as Infomap \cite{Rosvall:2008}) to the line graph, which encodes the path of the random walk on the line network under the Minimum Description Length (MDL) principle.

Line graph has been extended to clique graph \cite{evans:2010}, wherein cliques of a given order are represented as nodes in a weighted graph. The membership strength of a node $i$ to community $c$ is given by the fraction of cliques containing $i$ which are assigned to $c$. 

Although the link partitioning for overlapping detection seems conceptually natural, there is no guarantee that it provides higher quality detection than node based detection does \cite{Santo:2010} because these algorithms also rely on an ambiguous definition of community.

Note that a link-based extended modularity is also proposed by Nicosia in \cite{Nicosia:2009}. This measure is built on the belonging coefficients of \textit{links}. 
Let a link $l(i,j)$ connecting $i$ to $j$ for community $c$ be $\beta_{l(i,j),c}=F(a_{ic},a_{jc})$, then the expected belonging coefficient of any possible link $l(i,j)$ from node $i$ to a node $j$ in community $c$ can be defined as  $\beta_{l(i,j),c}^{out}=\frac{1}{|V|} \sum_{j\in V}F(a_{ic},a_{jc})$.
Accordingly, the expected belonging coefficient of any link $l(i,j)$ pointing to node $j$ in community $c$ is defined as $\beta_{l(i,j),c}^{in}=\frac{1}{|V|} \sum_{i\in V}F(a_{ic},a_{jc})$. The above belonging coefficients are used as weights for the probability of an observed link (first term in (\ref{eq:NicosiaQov})) and the probability of a link starting from $i$ to $j$ in the null model (second term in (\ref{eq:NicosiaQov})), respectively, resulting in the new modularity defined as 
\begin{equation}
\label{eq:NicosiaQov}
	Q_{ov}^{Ni}=\frac{1}{m} \sum_{c} \sum_{i,j\in V}\left[ \beta_{l(i,j),c}A_{i,j}-\beta_{l(i,j),c}^{out}\beta_{l(i,j),c}^{in}\frac{k_i^{out}k_j^{in}}{m}\right],
\end{equation}
where $k_i^{out(in)}$ is the number of outgoing (incoming) links of $i$ and $m$ is the total number of edges.
Note that $Q_{ov}^{Ni}$ depends on the link belonging coefficient $F(a_{ic},a_{jc})$, which could be the product, average, or maximum of $a_{ic}$ and $a_{jc}$.

\subsection{Local Expansion  and Optimization }
Algorithms utilizing local expansion and optimization are based on growing a \textit{natural} community \cite{LancichinettiNMI-LFM:2009} or a partial community. Most of them rely on a local benefit function that characterizes the quality of a densely connected group of nodes. 

Baumes \cite{Baumes1:2005} proposed a two-step process. First, the algorithm RankRemoval is used to rank nodes according to some criterion. Then the process iteratively removes highly ranked nodes until small, disjoint cluster cores are formed. These cores serve as seed communities for the second step of the process, Iterative Scan (IS), that expands the cores by adding or removing nodes until a local density function cannot be improved.  The proposed density function can be formally given as
$$f(c)=\frac{w_{in}^c}{w_{in}^c+w_{out}^c},$$ where $w_{in}^c$ and $w_{out}^c$ are the total internal and external weight of the community $c$. The worst-case running time is $O(n^2)$. The quality of discovered communities depends on the quality of seeds. Since the algorithm allows vertices to be removed during the expansion, IS has been shown to produce disconnected components in some cases. For this reason, a modified version called CIS was introduced in \cite{stephenRPI:2009}, wherein the connectedness is checked after each iteration. In the case that the community is broken into more than one part, only the one with the largest density is kept. CIS also develops a new fitness function
$$f(c)=\frac{w_{in}^c}{w_{in}^c+w_{out}^c}+\lambda e_p$$  incorporating the edge probability $e_p$. 
The parameter $\lambda$ controls how the algorithm behaves in sparse areas of the network. The addition of a node needs to strike a balance between the change in the internal degree density and the change in edge density. 

LFM \cite{LancichinettiNMI-LFM:2009} expands a community from a random seed node to form a natural community until the fitness function
\begin{equation}
\label{eq:lfm}
	f(c)=\frac{k_{in}^c}{(k_{in}^c+k_{out}^c)^\alpha}
\end{equation}
 is locally maximal, where $k_{in}^c$ and $k_{out}^c$ are the total internal and external degree of the community $c$, and $\alpha$ is the resolution parameter controlling the size of the communities. After finding one community, LFM randomly selects another node not yet assigned to any community to grow a new community. LFM depends significantly on the resolution parameter $\alpha$. The computational complexity for a fixed $\alpha$-value is roughly $O(n_cs^2)$, where $n_c$ is the number of communities and $s$ is the average size of communities. The worst-case complexity is $O(n^2)$. 

MONC \cite{MONC:2011} uses the modified fitness function of LFM
$$f(c)=\frac{k_{in}^c+1}{(k_{in}^c+k_{out}^c)^\alpha},$$
which allows a single node to be considered a community by itself.  This avoids violation of the principle of \textit{locality}. 
The proposed fitness function enables MONC to find the range of $\alpha$s (resolution parameter as in LFM) for which a set of nodes is locally optimal. Rather than numerical exploration of these $\alpha$ values, MONC calculates the next lowest value of $\alpha$ which results in further expansion and continues to expand the community. In the case that the natural community of a node $i$ is a subset of another node, the analysis of $i$ stops. In this way, MONC merges communities during processing and, as a result, uncovers the network faster than LFM.

OSLOM\footnote{\url{http://www.oslom.org.}} \cite{OSLOMLancichinetti:2011} tests the statistical significance of a cluster \cite{Bianconi:2008} with respect to a global null model (i.e., the random graph generated by the configuration model \cite{Molloy:1995}) during  community expansion. To grow the current community, the $r$ value is computed for each neighbor, which is the cumulative probability of having the number of internal connections equal or larger than the number of connections from a neighbor into this community in the null model. If the cumulative distribution of the smallest $r$ value is smaller than a given tolerance, it is considered to be significant, and the corresponding node is added to the community. Otherwise, the second smallest $r$ is checked and so on. OSLOM usually results in a significant number of outliers or singleton communities. The worst-case complexity in general is $O(n^2)$, while the exact complexity depends on the community structure of the underlying network being studied.

Rather than considering the original network, UEOC \cite{UEOCDiJin:2011} unfolds the community of a node based on the $l$-step transition probability of the random walk on the corresponding \textit{annealed} network  \cite{ANNNewman:2001}, which represents an ensemble of networks. After sorting nodes according to the transition probabilities in descending order, the natural community is extracted with some proper cutoff. The dominating time complexity is for calculating the transition matrix, which is $O(ln^2)$.  

OCA \cite{OCApadrol-sureda:2010} is based on the idea of mapping each node to a $d$-dimensional vector.  Each \textit{subset} of nodes $S$ is then defined as the sum of individual vectors in this set. The fitness function is defined as the directed Laplacian on function $O$, where $O$ is the squared Euclidean length of a subset vector. Like LFM, starting from some initial seeds, OCA tries to remove or add a node that results in the largest increase in the value of the fitness function. OCA requires finding the most negative eigenvalue of the adjacency matrix.

Chen \cite{DuanbingChen:2010} proposed selecting a node with maximal node strength based on two quantities $B(u,c)$  (called belonging degree) and the modified modularity $Q_{ov}$ for weighted networks. $Q_{ov}$ is defined as
\begin{equation}
\label{eq:ChenQov}
		Q_{ov}^C=\frac{1}{2m} \sum_{c}\sum_{i,j\in V}\left[A_{ij}-\frac{k_ik_j}{2m} \right]\beta_{ic}\beta_{jc},
\end{equation}
where
$\beta_{ic}=k_{ic}/\sum_{c'}k_{ic'}$ is the strength with which node $i$ belongs to community $c$, and $k_{ic}=\sum_{j\in c} w_{ij}$ is the total weight of links from $i$ into community $c$. $B(u,c)$ measures how tightly a node $u$ connects to a given community $c$ compared to the rest of the network. Given two thresholds $B^U$ and $B^L$, when expanding a community $c$, neighboring nodes with $B(u,c)$\textgreater$B^U$ are included in $c$. For nodes with $B^L\leq B(u,c) \leq B^U$, if $Q_{ov}$ increases after adding such a node, $u$ is added to $c$. 
The drawbacks of this algorithm are the rather arbitrary selection of the $B^U$ and $B^L$ thresholds and the expensive computation of $Q_{ov}$ whose complexity is $O(kn^2)$, where $k$ is the number of communities. 

iLCD\footnote{\url{http://cazabetremy.fr/Cazabet_remy/iLCD.html.}} \cite{iLCDCazabet:2010} is capable of detecting both static and temporal communities. Given a set of edges created at some time step, iLCD updates the existing communities by adding a new node if its number of second neighbors and number of robust second neighbors are greater than expected values. New edges are also allowed to create a new community if the minimum pattern is detected. Defining the similarity between two communities as the ratio of nodes in common, a merging procedure is performed to improve the detection quality if the similarity is high.  iLCD relies on two parameters for adding a node and merging two communities. The complexity of iLCD is $O(nk^2)$ in general, whose precise quantity depends on community structures and its parameters.

Seeds are very important for many local optimization algorithms. A clique has been shown to be a better alternative over an individual node as a seed, serving as the basis for a wide range of algorithms. EAGLE  
\cite{EAGLE:2009} uses the agglomerative framework to produce a dendrogram. First, all maximal cliques are found and made to be the initial communities. Then, the pair of communities with maximum similarity is merged. The optimal cut on the dendrogram is determined by the extended modularity with a weight based on the number of overlapping memberships in \cite{EAGLE:2009b}. Even without taking into account the time required to find all the maximal cliques, EAGLE is still computationally expensive with complexity $O(n^2+(h+n)s)$, where $s$ is the number of maximal cliques whose upper bound is $3^{n/3}$ (i.e., theoretically exponential) \cite{Cliques:1965}, and $h$ is the number of pairs of maximal cliques which are neighbors. This paper also defines an extended modularity that uses the number of communities to which a node belongs as a weight for $Q$ as 
\begin{equation}
\label{eq:ShenQov}
	Q_{ov}^E=\frac{1}{2m}\sum_c {\sum_{i,j \in c} {\left[ A_{ij}-\frac{k_ik_j}{2m}\right]{\frac{1}{O_iO_j}}}},
\end{equation}
where $O_{i}$ is the number of communities to which node $i$ belongs. 
This measure is in the same form as (\ref{eq:NepuszQov}), but with a coefficient defined based on the maximal clique. One may argue that they are identical as in \cite{SteveSurvey:2011}.

Similar to EAGLE, GCE\footnote{\url{https://sites.google.com/site/greedycliqueexpansion.}} \cite{GCE:2010} identifies maximum cliques as seed communities. It expands these seeds by greedily optimizing a local fitness function. GCE also removes communities that are similar to previously discovered using distance between communities $c_1$ and $c_2$ defined as 
$$1-\frac{|c_1\cap c_2|}{min(|c_1|,|c_2|)}.$$ If this distance is shorter than a parameter $\epsilon$, the communities are similar. The time complexity for greedy expansion is $O(mh)$, where $m$ is the number of edges, and $h$ is the number of cliques. 

In COCD \cite{DuNan:2008}, cores are a set of independent maximal cliques induced on each vertex. Two maximal cliques are said to be dependent if their \textit{closeness} function is positive. This function is a product of the differences between the size of internal links between two maximal cliques and the number of links connecting nodes appearing only in one of the two maximal cliques. Once the cores are identified, the remaining nodes are attached to cores with which they have maximum connections. COCD runs in $O(C_{max} \cdot Tri^2)$ in the worst case, where $C_{max}$ is the maximum size of the detected communities, and $Tri$ is the number of triangles, whose lower bound is $\frac{9mn-2n^3-2(n^2-3m)^{3/2}}{27}$ \cite{Triangle:1989} or $O(n^3)$ for a dense enough graph.

\subsection{Fuzzy Detection}
Fuzzy community detection algorithms quantify the strength of association between all pairs of nodes and communities.  In these algorithms, a soft membership vector, or belonging factor \cite{COPRA:2010}, is calculated for each node. A drawback of such algorithms is the need to determine the dimensionality $k$ of the membership vector.  This value can be either provided as a parameter to the algorithm or calculated from the data.

Nepusz \cite{Nepusz:2008} modeled the overlapping community detection as a nonlinear constrained optimization problem which can be solved by simulated annealing methods.  The objective function to minimize is
\begin{equation}
\label{eq:Nepusz}
	f=\sum_{i=1}^n{\sum_{j=1}^n{w_{ij}(\tilde{s_{ij}}-s_{ij})^2}},
\end{equation}
where $w_{ij}$ denotes the predefined weight, $\tilde{s_{ij}}$ is the \textit{prior} similarity between nodes $i$ and $j$, and the similarity $s_{ij}$ is defined as 
\begin{equation}
\label{eq:NepuszSim}
	s_{ij}=\sum_{c}{a_{ic}a_{jc}},
\end{equation}
where the variable $a_{ic}$ is the fuzzy membership of node $i$ in community $c$, subject to the total membership degree constraint in (\ref{eq:belonging}) and a non-empty community constraint. To determine the number of communities $k$, the authors increased the value of $k$ until the community structure does not improve as measured by a modified fuzzy modularity, which, by weighting $Q$ with the \textit{product} of a node's belonging factor, is defined as 
\begin{equation}
\label{eq:NepuszQov}
		Q_{ov}^{Ne}=\frac{1}{2m}\sum_c {\sum_{i,j \in c} {\left[A_{ij}-\frac{k_ik_j}{2m}\right]a_{ic}a_{jc}}},
\end{equation}
where $a_{ic}$ is the degree of membership of node $i$ in the community $c$.

Zhang \cite{ZhangFuzzy:2007} proposed an algorithm based on the spectral clustering framework  \cite{PhysRevE.74.036104,WhiteSmyth:2005}. Given an upper bound on the number of communities $k$, the top $k-1$ eigenvectors are computed. The network is then mapped into a $d$-dimensional Euclidean space, where $d \leq k-1$. Instead of  using $k$-means, fuzzy $c$-means (FCM) is used to obtain a soft assignment.  Both detection accuracy and computation efficiency rely on the user specified value $k$. With running time $O(mkh+nk^2h+k^3h)+O(nk^2)$, where $m$ is the number of edges, $n$ is the number of nodes, the first term is for the implicitly restarted Lanczos method, and the second term is for FCM, it is not scalable for large networks. An extended modularity that used the \textit{average} of the belonging factor is also proposed as 
$$Q_{ov}^Z=\sum_{c} {\left[ \frac{A(V'_c,V'_c)}{A(V,V)}-\left(\frac{A(V'_c,V)}{A(V,V)}\right)^2 \right]},$$
where $V'_c$ is the set of nodes in a community $c$, $w_{ij}$ is the weight of the link connecting nodes $i$ and $j$, $A(V'_c,V'_c)=\sum_{i,j\in V'_c}w_{ij}(a_{ic}+a_{jc})/2$, 
$A(V'_c,V)=A(V'_c,V'_c)+\sum_{i\in V'_c,j\in V \backslash V'_c}w_{ij}(a_{ic}+(1-a_{jc}))/2$, and
$A(V,V)=\sum_{i,j\in V}w_{ij}$.

Due to their probabilistic nature, mixture models provide an appropriate framework for overlapping community detection  \cite{MixtureNewMan:2007}. In general, the number of mixture models is equal to the number of communities, which needs to be specified in advance. In SPAEM\footnote{\url{http://www.code.google.com/p/spaem.}} \cite{SPAEMRenWei:2009},  the mixture model is viewed as a generative model for the links in the network. Suppose that $\pi_r$ is the probability of observing community $r$ and community $r$ selects node $i$ with probability $B_{r,i}$. For each $r$, $B_{r,i}$ is a multinomial across elements $i=1, 2, \cdots, n$, where $n$ is the number of nodes.  Therefore, $\sum_{i=1}^n {B_{r,i}} =1$. The edge probability $e_{ij}$ generated by such finite mixture model is given by
$$p(e_{ij}|\pi,B)=\sum_{r=1}^k{\pi_rB_{r,i}B_{r,j}}.$$
The total probability over all the edges present in the network is maximized by the Expectation-Maximization (EM) algorithm. As in \cite{Youngdokim:2011}, the optimal number of communities $k$ is identified based on the minimum description length. There is another algorithm called FOG\footnote{\url{http://www.casos.cs.cmu.edu/projects/ora.}}  \cite{DavisGeorge:2008} also trying to infer groups based on link evidence.

Similar mixture models can also be constructed as a generative model for nodes \cite{MMMQiangFu:2008}. In SSDE\footnote{\url{http://www.cs.rpi.edu/~purnej/code.php.}} \cite{SSDEJonathan:2011}, the network is first mapped into a $d$-dimensional space using the spectral clustering method. A Gaussian Mixture Model (GMM) is then trained via Expectation-Maximization algorithm. The number of communities is determined when the increase in log-likelihood of adding a cluster is not significantly higher than that of adding a cluster to random data which is uniform over the same space. 

Stochastic block model (SBM) \cite{SBMNowicki:2001} is another type of generative model for groups in the network. Fitting an empirical network to a SBM requires inferring model parameters  similar to GMM. In OSBM \cite{OSBMPierre:2011}, each node $i$ is associated with a latent vector (i.e., community assignment) $Z_i$ with $K$ independent Boolean variables $Z_{ik}\in\{0,1\}$, where $K$ is the number of communities, and  $Z_{ik}$ is drawn from a multivariate Bernoulli distribution. $Z$ is inferred by maximizing the posterior probability conditioned on the present of edges as in \cite{SPAEMRenWei:2009}. OSBM requires more efforts than mixture models because the factorization in the observed condition distribution for edges given $Z$ is in general intractable. MOSES\footnote{\url{http://sites.google.com/site/aaronmcdaid/moses.}} \cite{MOSES:2010} combines OSBM with the local optimization scheme, in which 
the fitness function is defined based on the observed condition distribution. MOSES greedily expands a community from edges. Unlike OSBM, no connection probability parameters are required as input. The worst-case time complexity is $O(en^2)$, where $e$ is the number of edges to be expanded. 

Non-negative Matrix Factorization (NMF) is a feature extraction and dimensionality reduction technique in machine learning that has been adapted to community detection. NMF approximately factorizes the feature matrix $V$ into two matrices with the non-negativity constraint as $V \approx WH$, where $V$ is $n \times m$, $W$ is $n \times k$ , $H$ is $k \times m$, and $k$ is the number of communities provided by users. $W$ represents the data in the reduced feature space. Each element $w_{i,j}$ in the normalized $W$ quantifies the dependence of node $i$ with respect to community $j$. In \cite{NMFZhangShihua:2007}, $V$ is replaced with the diffusion kernel, which is a function of the Laplacian of the network.  In \cite{MinaZarei-NMF:2009}, $V$ is defined as the correlation matrix of the columns of the Laplacian.  This results in better performance than \cite{NMFZhangShihua:2007}. In \cite{CNMFKunZhao:2010}, redundant constraints in the approximation are removed, reducing NMF to a problem of symmetrical non-negative matrix factorization (s-NMF). Psorakis \cite{BayesianNMFIoannis:2011} proposed a hybrid algorithm called Bayesian NMF\footnote{\url{http://www.robots.ox.ac.uk/~parg/software.html.}}. The matrix $V$, where each element $v_{ij}$ denotes a count of the interactions that took place between two nodes $i$ and $j$, is decomposed via NMF as part of the parameter inference for a generative model similar to OSBM and GMM. Traditionally, NMF is inefficient with respect to both time and memory constraints due to the matrix multiplication. In the version of  \cite{BayesianNMFIoannis:2011}, the worst-case time complexity is $O(kn^2)$, where $k$ denotes the number of communities.

Wang et al. \cite{WangLineGraph3:2009} combined disjoint detection methods with local optimization algorithms. First, a partition is obtained from any algorithm for disjoint community detection. Communities attempt to add or remove nodes. The difference, called variance, of two fitness scores on a community, either including a node $i$ or removing node $i$, is computed. The normalized variances form a fuzzy membership vector of node $i$.

Ding \cite{AffinityFanDing:2010} employed the affinity propagation clustering algorithm  \cite{affinitypropagation:2007} for overlapping detection, in which clusters are identified  by representative exemplars. First, nodes are mapped as data points in the Euclidean space via the commute time kernel (a function of the inverse Laplacian). The similarity between nodes is then measured by the cosine distance. Affinity propagation reinforces two types of messages associated with each node, the responsibility $r(i,k)$ and the availability $a(i,k)$. The probability for assigning node $i$ into the cluster represented by exemplar node $k$ is computed by equation $p(i,k)=e^{\hat{r}(i,k)}$, where $\hat{r}$ is the normalized responsibility as in \cite{Geweniger:2009}.

\subsection{Agent-Based and Dynamical Algorithms}
The label propagation algorithm \cite{Raghavan:2007,JieruiXieLPA:2010}, in which nodes with same label form a community, has been extended to overlapping community detection by allowing a node to have multiple labels. In COPRA\footnote{\url{http://www.cs.bris.ac.uk/~steve/networks/software/copra.html.}} \cite{COPRA:2010}, each node updates its belonging coefficients by averaging the coefficients from all its neighbors at each time step in a synchronous fashion. The parameter $v$ is used to control the maximum number of communities with which a node can associate. The time complexity is $O(vm \log (vm/n))$ per iteration. 

SLPA\footnote{\url{https://sites.google.com/site/communitydetectionslpa.}} \cite{JieruiXieSLPA-ICDM:2011,JieruiXieSLPA-pkdd:2012} is a general speaker-listener based information propagation process. It spreads labels between nodes according to pairwise interaction rules. Unlike \cite{Raghavan:2007,COPRA:2010}, where a node forgets knowledge gained in the previous iterations, SLPA provides each node with a memory to store received information (i.e., labels). The probability of observing a label in a node's memory is interpreted as the membership strength. SLPA does not require any knowledge about the number of communities, which is determined by the clustering of labels in the network. The time complexity is $O(tm)$, linear in the number of edges $m$, where $t$ is a predefined maximum number of iterations (e.g., $t\geq 20$). SLPA can also be adapted for weighted and directed networks by generalizing the interaction rules, known as SLPAw.

A game-theoretic framework is proposed in \cite{ChenWei:2010}, in which a community is associated with a Nash local equilibrium. A gain function and a loss function are associated with each agent. The game assumes that each agent is selfish and selects to join, leave and switch communities based on its own utility.  An agent is allowed to joint multiple communities to handle overlapping, so long as it results in increased utility. The time complexity to find the best local operation for an agent $i$ is $O(|L_i|\cdot|L(N_i)|\cdot k_i)$, where $L_i$  is the communities that agent $i$ wants to joint, $L(N_i)$ is the set of communities that $i$'s neighbors want to joint, and $k_i$ is the node degree. The time takes to reach a local equilibrium is bounded by $O(m^2)$, where $m$ is the number of edges.

A process in which particles walk and compete with each other to occupy nodes is presented in  \cite{BreveFabricio:2009}. Particles represent different communities. Each node has an instantaneous ownership vector (similar to belonging factor) and a long term ownership vector. At each iteration, each particle takes either a random walk or a deterministic walk to one of its neighbors with some probability. If the random walk is performed, the visited neighbor updates its instantaneous ownership vector; otherwise, the long term ownership vector is updated. At the end of the process, the long term ownership vector is normalized to produce a soft assignment. Different from SLPA and COPRA, this algorithm takes a semi-supervised approach. It requires at least one labeled node per class.

Multi-state spin models \cite{Reichardt:2004,Lu:2009}, in which a spin is assigned to each node, can also be applied to community detection. One of such models is $q$-state Potts model \cite{Blatt:1996,Reichardt:2004}, where $q$ is the number of states that a spin may take, indicating the maximum number of communities. The community detection problem is equivalent to the problem of minimizing the Hamiltonian of the model. In the ground states (i.e., local minima of the Hamiltonian), the set of nodes with the same spin state form a community. The overlap of communities is linked to the degeneracy of the minima of the Hamiltonian \cite{Reichardt:2006}.  Although a co-appearance matrix keeps track of how frequently nodes $i$ and $j$ have been grouped together over multiple runs, it is not clear how to aggregate this information into overlapping communities when analyzing large networks. Another Potts model-like approach was proposed in \cite{Peter:2009} to evaluate the hierarchical or multiresolution structure of a graph via information-based replica correlations.

Synchronization of a system that consists of coupled phase oscillators is able to uncover community structures. In such a model (e.g., the Kuramoto model) the phase of each unit evolves in time according to the predefined dynamics. The set of nodes with the same phase or frequency can be viewed as a community \cite{AlexSyn:2006} while nodes that do not match any observed dynamic behaviors can be considered overlapping nodes \cite{synDataLi:2008}. Like methods utilizing a Potts model, such algorithms are parameter dependent.

\subsection{Others}

CONGA\footnote{\url{http://www.cs.bris.ac.uk/~steve/networks/software/conga.html.}} \cite{CONGA:2007} extends Girvan and Newman's divisive clustering algorithm (GN) \cite{GirNew02} by allowing a node to split into multiple copies. Both \textit{splitting betweenness}, defined by the number of shortest paths on the imaginary edge, and the conventional edge betweenness are considered. CONGA inherits the high computational complexity of GN.  In a more refined version, CONGO  \cite{CONGO:2008} uses local betweenness to optimize the speed. Gregory \cite{SteveDisjoint:2009} also proposed to perform disjoint detection algorithms on the network produced by splitting the node into multiple copies using the split betweenness. 

Zhang\footnote{\url{http://dbgroup.cs.tsinghua.edu.cn/zhangyz/kdd09.}} \cite{YuzhouZhang:2009} proposed an iterative process that reinforces the network topology and \textit{propinquity} that is interpreted as the probability of a pair of nodes belonging to the same community. The propinquity between two vertices is defined as the sum of the number of direct links, number of common neighbors and the number of links within the common neighborhood. Given the topology, propinquity is computed.  Propinquity above a certain threshold is then used to redistribute links, updating the topology. If the propinquity is large, a link is added to the network; otherwise, the link is removed. The propinquity can be used to perform micro clustering on each vertex to allow overlap. 

Kov\'acs et al. \cite{Landscapespalotai:2010} proposed an approach focusing on centrality-based influence functions. Community structures are interpreted as hills of the influence landscape. For each node $i$, the influence over each link $f_i(j,k)$ is computed.  Links within a community should have higher influence than those linking distant areas of the network. The influence on a given link $c(j,k)$ is the sum of $f_i(j,k)$ over all nodes. The function $c(j,k)$ over each link defines the community landscape, wherein the communities are determined by local maxima and their surrounding regions.

Rees \cite{Rees:2010} proposed an algorithm to extract the overlapping communities from the \textit{egonet}, which is a subgraph including a center node, its neighbors, and the links around them.  When all egonets are induced, each center is removed, creating small connected components among neighbors. Then, the center node is added back to each of these components to form so-called \textit{friendship group}. Clearly, each center node can be in multiple friendship groups. The overlapping communities are determined by merging all  friendship groups. 

Inspired by OPTIC \cite{OPTICS:1999}, an algorithm based on techniques from visualization was proposed in  \cite{ONDOCS:2009}.  Nodes are ordered according to the reachability score (RS) with respect to a starting node. The reachability is based on the probability of the existence of a link between two nodes. By scanning through the obtained \textit{sequence} of nodes, a community containing \textit{consecutive} nodes with RS larger than a \textit{community threshold} is found. Clearly, this algorithm is hard to apply to large networks and requires the introduction of a community threshold.

\section{Evaluation Criteria}
\label{sec:cr}
Evaluating the quality of a detected partitioning or cover is nontrivial, and extending evaluation measures from disjoint to overlapping communities is rarely straightforward. Unlike disjoint community detection, where a number of measures have been proposed for comparing \textit{identified} partitions with the \textit{known} partitions \cite{Danon05comparingcommunity,wwwLeskovec:2010}, only a few measures are suitable for a set of overlapping communities. Two most widely used measures are the Normalized Mutual Information (NMI) and Omega Index.

\subsection{Normalized Mutual Information}
Lancichinetti \cite{LancichinettiNMI-LFM:2009} has extended the notion of normalized mutual information to account for overlap between communities. For each node $i$ in cover $C'$, its community membership can be expressed as a binary vector of length $|C'|$  (i.e., the number of clusters in $C'$).  $(x_i)_k=1$ if node $i$ belongs to the $k^{th}$ cluster $C'_k$;  $(x_i)_k=0$ otherwise.  The $k^{th}$ entry of this vector can be viewed as a random variable $X_k$, whose probability distribution is given by $P(X_k=1)=n_k/n$, $P(X_k=0)=1-P(X_k=1)$, where $n_k=|C'_k|$ is the number of nodes in the cluster $C'_k$ and $n$ is the total number of nodes. The same holds for the random variable $Y_l$ associated with the $l^{th}$ cluster in cover $C''$. Both the empirical marginal probability distribution $P(X_k)$ and the joint probability distribution $P(X_k,Y_l)$ are used to further define entropy $H(X)$ and $H(X_k,Y_l)$.

The conditional entropy of a cluster $X_k$ given $Y_l$ is defined as $H(X_k|Y_l)=H(X_k,Y_l)-H(Y_l)$. The entropy of $X_k$ with respect to the entire vector $Y$ is based on the best matching between $X_k$ and any component of Y given by 
$$H(X_k|Y)=min_{l\in \{1, 2, \cdots, |C''| \}} H(X_k|Y_l).$$
The normalized conditional entropy of a cover $X$ with respect to $Y$ is 
$$H(X|Y)=\frac{1}{|C'|} \sum_k {\frac{H(X_k|Y)}{H(X_k)}}.$$
In the same way, one can define $H(Y|X)$. Finally the NMI for two covers $C'$ and $C''$ is given by
\begin{equation}
\label{eq:nmi}
	NMI(X|Y)=1-\left[ H(X|Y)+ H(Y|X) \right]/2.
\end{equation}
The extended NMI is between 0 and 1, with 1 corresponding to a perfect matching. Note that this modified NMI does not reduce to the standard formulation of NMI when there is no overlap.
 
\subsection{Omega Index}
\textit{Omega Index} \cite{Omega:1988} is the overlapping version of the Adjusted Rand Index (ARI) \cite{Hubert:1985}. It is based on \textit{pairs} of nodes in $agreement$ in two covers. Here, a pair of nodes is considered to be in \textit{agreement} if they are clustered in \textit{exactly} the same number of communities (possibly none). That is, the omega index considers how many pairs of nodes belong together in no clusters, how many are placed together in exactly one cluster, how many are placed in exactly two clusters, and so on.

Let $K_1$ and $K_2$ be the number of communities in covers $C_1$ and $C_2$, respectively, the omega index is defined as \cite{SteveSurvey:2011,MONC:2011}
\begin{equation}
\label{eq:omega}
		\omega(C_1,C_2)=\frac{\omega_u(C_1,C_2)-\omega_e(C_1,C_2)}{1-\omega_e(C_1,C_2)}.
\end{equation}
The unadjusted omega index $\omega_u$ is defined as
$$\omega_u(C_1,C_2)=\frac{1}{M} \sum_{j=0}^{max(K_1,K_2)} {|t_j(C_1)\cap t_j(C_2)|},$$ where $M$ equal to $n(n-1)/2$ represents the number of node pairs and $t_j(C)$ is the set of pairs that appear exactly $j$ times in a cover $C$.
The expected omega index in the null model $\omega_e$ is given by
$$\omega_e(C_1,C_2)=\frac{1}{M^2} \sum_{j=0}^{max(K_1,K_2)} {|t_j(C_1)|\cdot| t_j(C_2)|}.$$

The subtraction of the expected value in (\ref{eq:omega}) takes into account agreements resulting from chance alone. The larger the omega index is, the better the matching is between two covers. A value of 1 indicates perfect matching. When there is no overlap, the omega index reduces to the ARI.

In addition to NMI and Omega, some other measures have been proposed, such as the generalized external indexes  \cite{FuzzyExternInd:2007,FuzzyExternInd:2010} and the fuzzy rand index \cite{FuzzyRIndex:2009}.

\section{Benchmarks}
It is necessary to have good benchmarks to both study the behavior of a proposed community detection algorithm and to compare the performance across various algorithms.  In order to accurately perform these two analyses, networks in which the ground truth is known are needed.  This requirement implies that real-world networks, which are often collected from online or observed interactions, do not paint a clear enough picture due to their lack of ``ground truth".  In light of this requirement, we begin our discussion with synthetic networks. In the GN benchmark  \cite{GirNew02}, equal size communities are embedded into a network for a given expected degree and a given mixing parameter $\mu$ that measures the ratio of internal connections to outgoing connections.  One drawback of this benchmark is that it fails to account for the heterogeneity in complex networks. Another is that it does not allow embedded communities to overlap. A few benchmarks have been proposed for testing overlapping community detection algorithms, all of which are special cases of the planted \textit{l}-partition model \cite{plantedLpartition:2001} just like GN.

Sawardecker \cite{Sawardecker:2009} proposed an extension of GN, in which the probability $p_{ij}$ of an edge being present in the network is a non-decreasing function based solely on  the set of co-memberships of nodes $i$ and $j$. With the definition $p_{ij}=p_k$, parameter $p_k$ is the connection probability of nodes $i$ and $j$ that co-occur $k$ times, subject to $p_0 < p_1 \leq p_2 \leq \cdots$. 

The LFR\footnote{\url{http://sites.google.com/site/andrealancichinetti/files.}} benchmark proposed in  \cite{LFR:2008} introduces heterogeneity into degree and community size distributions of a network. These distributions are governed by power laws with exponents $\tau_1$ and $\tau_2$, respectively. To generate overlapping communities, $O_n$, the fraction of overlapping nodes is specified and each node is assigned to $O_m\geq 1$ communities. The generating procedure is equivalent to generating a bipartite network where the two classes are the communities and nodes subject to the requirement that the sum of community sizes equals the sum of node memberships. LFR also provides a rich set of parameters to control the network topology, including the mixing parameter $\mu$, the average degree $\overline{k}$, the maximum degree $k_{max}$,  the maximum community size $c_{max}$, and the minimum community size $c_{min}$.

The LFR model brings benchmarks closer to the features observed in real-world networks. However, requiring that overlapping nodes interact with the same number of embedded communities is unrealistic in practice. A simple generalization, where each overlapping node may belong to different number of communities has been considered in  \cite{MOSES:2010}. 

In \cite{SteveSurvey:2011}, crisp communities from LFR are converted to fuzzy associations by adding a belonging coefficient to the occurrence of nodes.  This coefficient can be defined as 
				$$p_{ij}=s_{ij}p_1+(1-s_{ij})p_0,$$
where $p_k$ is the same as in Sawardecker's model and $s_{ij}=\sum_{c\in C} \alpha_{ic}\alpha_{jc}$ is the similarity of node $i$ and $j$ as defined in (\ref{eq:Nepusz}). In other words, the probability of an edge being present depends not only on the number of communities in which nodes $i$ and $j$ appear together but also on their degree of belonging to these communities.
\begin{table*}[tbp]
\centering
\caption{Algorithms included in the experiments.}
\label{table:algs}
\scalebox{0.8} {
	\addtolength{\tabcolsep}{-0pt}  
\begin{tabular}{cccc} \hline
\textbf{Algorithm} & \textbf{Reference} & \textbf{Complexity} & \textbf{Imp} \\ \hline
 CFinder & \cite{CPM:2005} & - & C++ \\ 
 LFM & \cite{LancichinettiNMI-LFM:2009} & $O(n^2)$ & C++ \\ 
 EAGLE & \cite{EAGLE:2009} & $O(n^2+(h+n)s)$ & C++ \\ 
 CIS & \cite{stephenRPI:2009} &  $O(n^2)$ & C++ \\ 
 GCE & \cite{GCE:2010} & $O(mh)$ & C++ \\ 
 COPRA & \cite{COPRA:2010} & $O(vm \log (vm/n))$ & Java \\ 
 Game & \cite{ChenWei:2010} & $O(m^2)$ & C++ \\ 
 NMF & \cite{BayesianNMFIoannis:2011} & $O(kn^2)$ & Matlab \\
 MOSES & \cite{MOSES:2010} & $O(en^2)$ & C++ \\ 
 Link & \cite{YYLinkClustering:2010} & $O(nk_{max}^2)$ & C++ \\ 
 iLCD & \cite{iLCDCazabet:2010} & $O(nk^2)$ & Java \\ 
 UEOC & \cite{UEOCDiJin:2011} & $O(ln^2)$ & Matlab \\ 
 OSLOM & \cite{OSLOMLancichinetti:2011} & $O(n^2)$ & C++ \\ 
 SLPA & \cite{JieruiXieSLPA-ICDM:2011,JieruiXieSLPA-pkdd:2012} & $O(tm)$ & C++\\ \hline 
\end{tabular}
}
\end{table*}
\section{Tests on Synthetic Networks}

In this section, we empirically compare the performance of different algorithms on LFR networks.
We focus on algorithms which produce a crisp assignment of vertices to communities. In total, 14 algorithms were collected and tested.  They are listed in Table \ref{table:algs}. Note that the time complexity given is for the worst case. 

For algorithms with tunable parameters, the results with the best setting are reported. For LFM, we varied $\alpha$ from 0.8 to 1.6 with an interval 0.1, within which good results have previously been reported  \cite{LancichinettiNMI-LFM:2009,GCE:2010}. For iLCD, $fRatio$ is from $\{0.75, 0.5, 0.35\}$ and $bThreshold$ is from $\{0.5, 0.3, 0.2\}$ as suggested by the authors. For GCE, the minimum clique size $k$ ranges from 3 to 8. For CFinder, $k$ ranges from 3 to 8. For OSLOM, we considered the first two levels. For Link, the threshold varies from 0.1 to 0.9 with an interval 0.1. For COPRA, parameter $v$ is taken from the range [1,10]. For SLPA, parameter $r$ varies from 0.05 to 0.5 with an interval 0.05. Since COPRA and SLPA are non-deterministic, we repeated each of them 10 times on each network instantiation. For NMF, which returns a fuzzy assignment, we applied the same threshold as SLPA to convert it to a crisp assignment. 

For each parameter set generated via LFR, we generated 10 instantiations. We used networks with sizes $n\in\{1000,5000\}$. The average degree is kept at $\overline{k}=10$, which is of the same order as most large real-world social networks\footnote{\url{snap.stanford.edu/data.}}. 
The rest of the parameters of LFR generator are set similar to those in \cite{LancichinettiComp:2009}: 
node degrees and community sizes are governed by power law distributions with exponents $\tau_1=2$ and $\tau_2=1$ respectively, the maximum degree is $k_{max}=50$, and community sizes vary in both small range $s=(10,50)$ and large range $b=(20,100)$. The mixing parameter $\mu$ is from $\{0.1, 0.3\}$, which is the expected fraction of links through which a node connects to other nodes in the same community.

The degree of overlap is determined by two parameters. $O_n$ is the number of overlapping nodes, and $O_m$ is the number of communities to which each overlapping node belongs. $O_n$ is set to 10\% and 50\% of the total number of nodes, indicating low and high \textit{overlapping density} respectively. Instead of fixing $O_m$ \cite{LancichinettiComp:2009,COPRA:2010}, we also allow it to vary from 2 to 8 indicating the \textit{overlapping diversity} of overlapping nodes. By increasing the value of $O_m$, we create harder detection tasks. This also allow us to look in greater detail at the detection accuracy at the node level. 
 
Two previously discussed measures, Omega and NMI, are used to quantify the quality of the cover discovered by an algorithm. 

\subsection{Effects of $\mu$, $n$ and $O_m$}
\begin{figure}[tp]	
	\centering
    \includegraphics[width=1.0\linewidth]{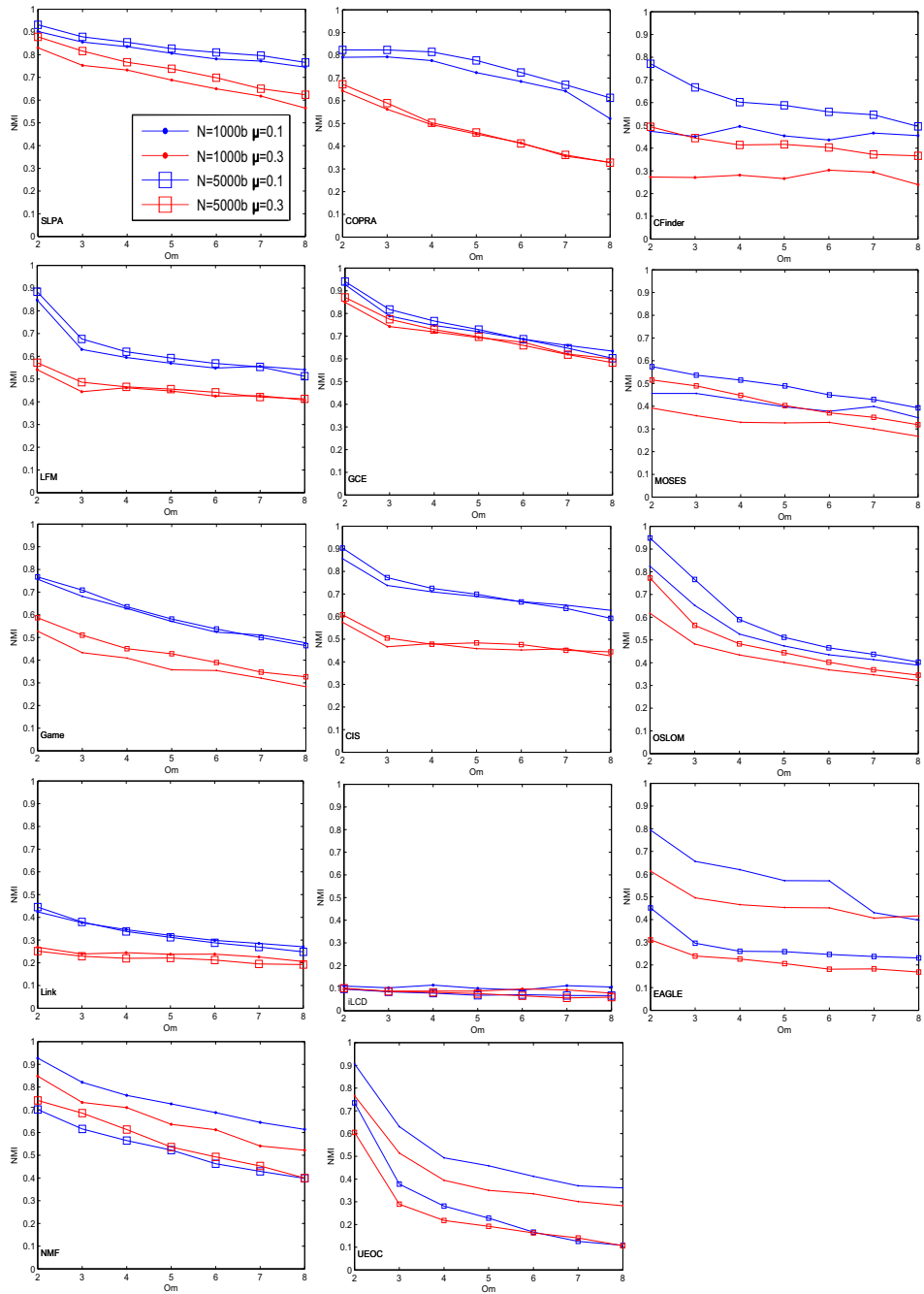} %
		\caption{The effects of network size $n$ and mixing parameter $\mu$ on LFR networks. Plots show NMI's for networks with large community size range and $O_n=10\%$. The order of subplots is from general behaviors to exceptions (see text).}
	 	\label{fig:Effect-N-Mu-NMI-bn01-all-3col}
\end{figure}

We first examined how the performance, measured by NMI, changes as the number of memberships $O_m$ varies from small to large values (i.e., 2 to 8) for different network sizes ($n$) and intra community strength ($\mu$) in Figure \ref{fig:Effect-N-Mu-NMI-bn01-all-3col}.  

In general, changes in the network topology, especially the mixing value $\mu$, have a similar impact for disjoint community detection. That is, the larger the value of $\mu$, the poorer the results produced by detection algorithms (i.e., red curve $<$ blue curve in Figure \ref{fig:Effect-N-Mu-NMI-bn01-all-3col}) due to the fact that the connection inside communities is weak for larger $\mu$ . This is true for the majority of algorithms with the only exception NMF (see the 5000b case). On the other hand, increasing network size from 1000 to 5000 typically results in slightly better performance (i.e., square $>$ dot in Figure \ref{fig:Effect-N-Mu-NMI-bn01-all-3col}), with prominent exceptions for EAGLE, NMF and UEOC. Slight fluctuations in performance are observed for iLCD and Link. Detection performance typically decays at a moderate rate as the diversity of overlapping increases (i.e., $O_m$ getting larger), except for OSLOM and UEOC. 
 		 	
\begin{figure}[tp]	
	\centering
    \includegraphics[width=1.0\linewidth]{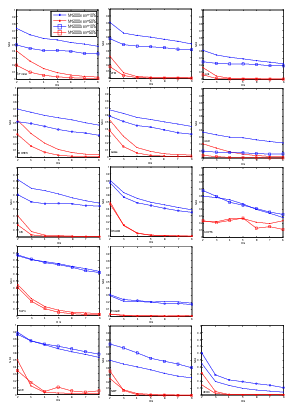} %
		\caption{The effects of community size range and overlapping density $O_n$ on LFR networks. Plots show NMI's for networks with $n=5000$ and $\mu=0.3$. The order of subplots is from general behaviors to exceptions (see text).}
	 	\label{fig:Effect-cs-On-NMI-N5000Mu03-all-3col}
\end{figure} 	
	 	
\subsection{Effects of Community Size Range and Overlapping Density $O_n$}
We evaluated the effects of $O_n$ and community size ranges on each individual algorithm on networks with $n=5000$ and $\mu=0.3$. Results for NMI are shown in Figure \ref{fig:Effect-cs-On-NMI-N5000Mu03-all-3col}.

As expected, the performance of detection consistently and significantly drops in the case there are many overlapping nodes for all algorithms (i.e., red curves ($O_n=50\%$) $<$ blue curves ($O_n=10\%$)). However, the difference in performance between small and large community size ranges (gaps between two curves with the same color) is more prominent in the case of low overlapping density. 
		 	
Interestingly, the NMI's for networks with small community size $s=(10,50)$ are typically higher than those for networks with large community size range $b=(20,100)$ (i.e., dot $>$ square in Figure \ref{fig:Effect-cs-On-NMI-N5000Mu03-all-3col}). It appears that the well known resolution limit does not play a role here since all the tested algorithms are neither based on a modularity nor an extended modularity. This is evidenced by algorithms including CFinder, LFM, Link, MOSES, Game, iLCD, CIS and OSLOM that have a significant performance gap between small and large community size ranges. Given only small variances in performance in two tested ranges, we also conclude that the community size range has limited impact on algorithms including SLPA, COPRA, EAGLE and GCE.
 	 
\begin{figure}[tp]	
	\centering
    \includegraphics[width=1.0\linewidth]{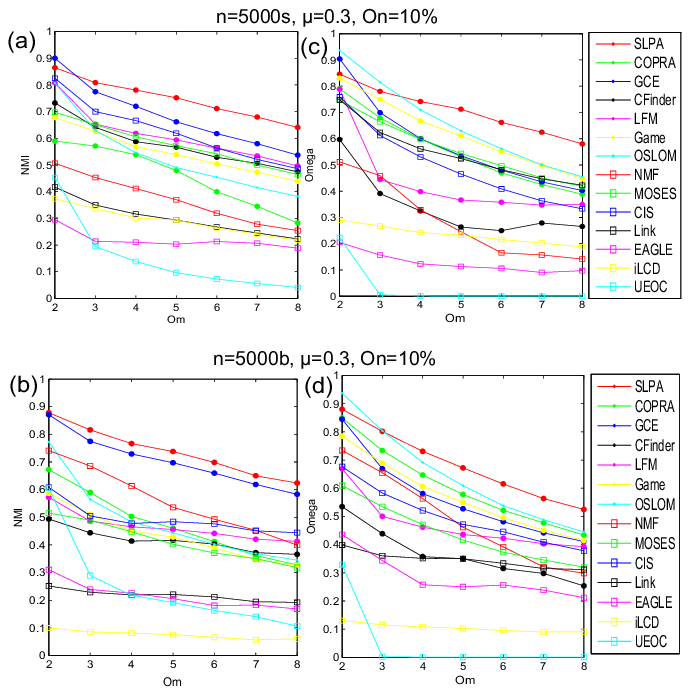} %
		\caption{Evaluations of overlapping community detection on LFR networks with low overlap density $O_n=10\%$. Left column: NMI as a function of the number of memberships $O_m$; Right column: Omega as a function of the number of memberships $O_m$. Results for small community size range are shown in top row (i.e., (a) and (c)), and results for large community size range are shown in bottom row (i.e., (b) and (d)). All resutls are from networks with $n=5000$ and $\mu=0.3$.}
	 	\label{fig:NMI-OMG-N5000sbMu03On01-col}
\end{figure}

\begin{figure}[tp]	
	\centering
    \includegraphics[width=1.0\linewidth]{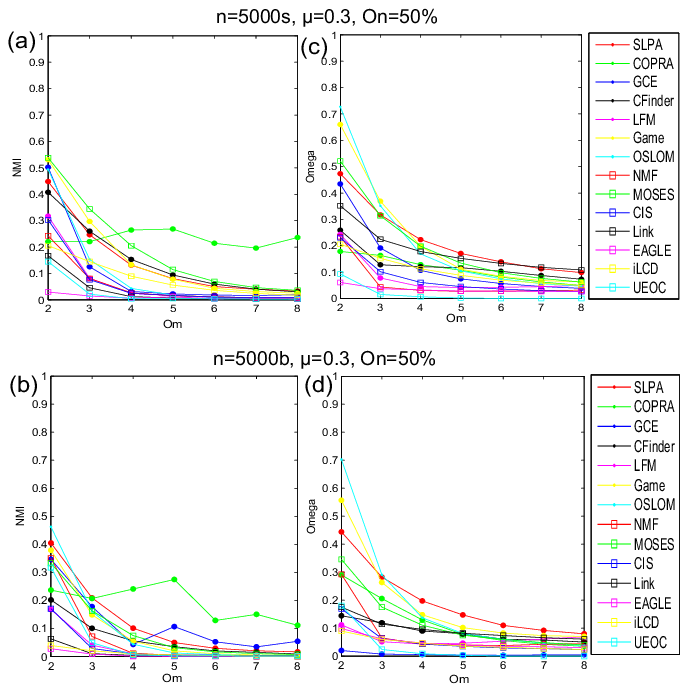} %
		\caption{Evaluations of overlapping community detection on LFR networks with high overlap density $O_n=50\%$. Left column: NMI as a function of the number of memberships $O_m$; Right column: Omega as a function of the number of memberships $O_m$. Results for small community size range are shown in top row (i.e., (a) and (c)), and results for large community size range are shown in bottom row (i.e., (b) and (d)). All resutls are from networks with $n=5000$ and $\mu=0.3$.}
	 	\label{fig:NMI-OMG-N5000sbMu03On05-col}
\end{figure}

\subsection{Ranking for Community Detection} 
Extensive comparisons have been conducted over different overlapping densities and community size ranges. Performance measured by NMI and Omega for $n=5000$ and $\mu=0.3$ is shown in Figures \ref{fig:NMI-OMG-N5000sbMu03On01-col} and \ref{fig:NMI-OMG-N5000sbMu03On05-col}.

To provide an intuitive way for both comparing two measures and also summarizing the vast volume of experiment results, we propose $RS_M(i)$, the averaged ranking score for a given algorithm $i$ with respect to some measure $M$ as follows:
\begin{equation}
\label{eq:ranking}
		RS_M(i)=\sum_{j=1} rank(i,O_m^j),
\end{equation}
where $O_m^j$ is the number of memberships (diversity) in $\{2, 3, \cdots, 8\}$, and function \textit{rank} returns the ranking of algorithm $i$ for the given $O_m$. Sorting  $RS_M$ in increasing order gives the final ranking among algorithms. Whenever it is clear from context, we use the term \textit{ranking} to refer to the final rank without the actual score value.

\begin{table*}[tp]
\centering
\caption{The community detection ranking for $n=5000$, $\mu=0.3$ and low overlapping density $O_n=10\%$.}
\label{table:rankingOn01}
\scalebox{0.8} {
	\addtolength{\tabcolsep}{-1pt}  
\begin{tabular}{c||cccc||cc} \hline
\textbf{Rank}	& \textbf{$RS^s_{NMI}$} & \textbf{$RS^s_{Omega}$}  & \textbf{$RS^b_{NMI}$} & \textbf{$RS^b_{Omega}$}	& \textbf{$RS^*_{NMI,Omega}$}	& \textbf{$RS^*_{F}$}  \\ \hline
1&SLPA&SLPA&SLPA&SLPA&SLPA&SLPA\\ 
2&GCE&OSLOM&GCE&OSLOM&GCE&CFinder\\ 
3&CIS&Game&NMF&COPRA&OSLOM&Game\\ 
4&LFM&GCE&CIS&Game&CIS&OSLOM\\ 
5&MOSES&MOSES&COPRA&GCE&Game&MOSES\\ 
6&CFinder&COPRA&OSLOM&CIS&COPRA&COPRA\\ 
7&Game&Link&LFM&NMF&LFM&iLCD\\ 
8&OSLOM&CIS&Game&LFM&MOSES&Link\\ 
9&COPRA&LFM&CFinder&MOSES&NMF&LFM\\ 
10&NMF&CFinder&MOSES&CFinder&CFinder&UEOC\\ 
11&Link&NMF&Link&Link&Link&EAGLE\\ 
12&iLCD&iLCD&EAGLE&EAGLE&EAGLE&GCE\\ 
13&EAGLE&EAGLE&UEOC&iLCD&iLCD&CIS\\ 
14&UEOC&UEOC&iLCD&UEOC&UEOC&NMF\\ \hline
\end{tabular}
}
\end{table*}

\begin{table*}[tp]
\centering
\caption{The community detection ranking for $n=5000$, $\mu=0.3$ and high overlapping density $O_n=50\%$.}
\label{table:rankingOn05}
\scalebox{0.8} {
	\addtolength{\tabcolsep}{-1pt}  
\begin{tabular}{c||cccc||cc} \hline
\textbf{Rank}	& \textbf{$RS^s_{NMI}$} & \textbf{$RS^s_{Omega}$}  & \textbf{$RS^b_{NMI}$} & \textbf{$RS^b_{Omega}$}	& \textbf{$RS^*_{NMI,Omega}$}	& \textbf{$RS^*_{F}$}  \\ \hline
1&MOSES&SLPA&COPRA&SLPA&SLPA&Link\\ 
2&COPRA&Link&SLPA&Game&MOSES&UEOC\\ 
3&CFinder&Game&GCE&OSLOM&Game&SLPA\\ 
4&Game&MOSES&MOSES&Link&COPRA&Game\\ 
5&SLPA&CFinder&CFinder&MOSES&CFinder&LFM\\ 
6&GCE&OSLOM&Game&CFinder&OSLOM&CFinder\\ 
7&iLCD&COPRA&OSLOM&COPRA&Link&CIS\\ 
8&OSLOM&GCE&LFM&LFM&GCE&MOSES\\ 
9&CIS&iLCD&CIS&NMF&LFM&OSLOM\\ 
10&LFM&LFM&NMF&EAGLE&CIS&iLCD\\ 
11&NMF&CIS&UEOC&CIS&iLCD&GCE\\ 
12&Link&NMF&iLCD&iLCD&NMF&COPRA\\ 
13&EAGLE&EAGLE&Link&UEOC&EAGLE&EAGLE\\ 
14&UEOC&UEOC&EAGLE&GCE&UEOC&NMF\\ \hline
\end{tabular}
}
\end{table*} 	 	

The results for low overlapping density case in Figure \ref{fig:NMI-OMG-N5000sbMu03On01-col} are summarized as four rankings including $RS^s_{NMI}$, $RS^b_{NMI}$, $RS^s_{Omega}$ and $RS^b_{Omega}$ in Table \ref{table:rankingOn01}, where $RS^{s(b)}_{NMI(Omega)}$ denotes the ranking based on NMI (or Omega) for networks with small (or large) community size range. Results for high overlapping density case in Figure \ref{fig:NMI-OMG-N5000sbMu03On05-col} are summarized in Table \ref{table:rankingOn05}. 

We first compared pairwise similarities of ($RS^s_{NMI}$, $RS^s_{Omega}$) and ($RS^b_{NMI}$, $RS^b_{Omega}$). As shown, among the top seven (half of the total fourteen) algorithms in two rankings, there are 4 pairs of matches (ignoring the exact order) for $O_n=10\%$ and 5 pairs of matches for $O_n=50\%$ for ($RS^s_{NMI}$, $RS^s_{Omega}$). Even more, there are 6 pairs of matches for both $O_n=10\%$ and $O_n=50\%$ for ($RS^b_{NMI}$, $RS^b_{Omega}$). This suggests that NMI and Omega provide similar overall evaluation to some extent.

Based on these four rankings, we further derive the average ranking $RS^*_{NMI,Omega}$ as the overall community detection performance. In this final ranking, the top seven  algorithms are exclusively agent-based algorithms (SLPA, Game and COPRA) and local expansion based algorithms (GCE, OSLOM, CIS, and LFM), which significantly outperform the others for networks with low overlapping density (see Figure \ref{fig:NMI-OMG-N5000sbMu03On01-col}). 

For high overlapping density, agent-based algorithms remain in the top seven, together with MOSES representing fuzzy algorithms, CFinder representing clique algorithms, Link representing link partitioning. However, given the fact that the performance is actually fairly low (most of them are less than 0.5 for $O_m>2$) shown in Figure \ref{fig:NMI-OMG-N5000sbMu03On05-col}, it is fair to conclude that all the algorithms do not yet achieve satisfying performance for networks with high overlapping density and high overlapping diversity (e.g., for $O_n=50\%$ and $O_m>2$).

\begin{figure*}[tp]	
\centering
	\begin{minipage}[t]{0.45\linewidth}
	\centering
		\includegraphics[scale=0.56]{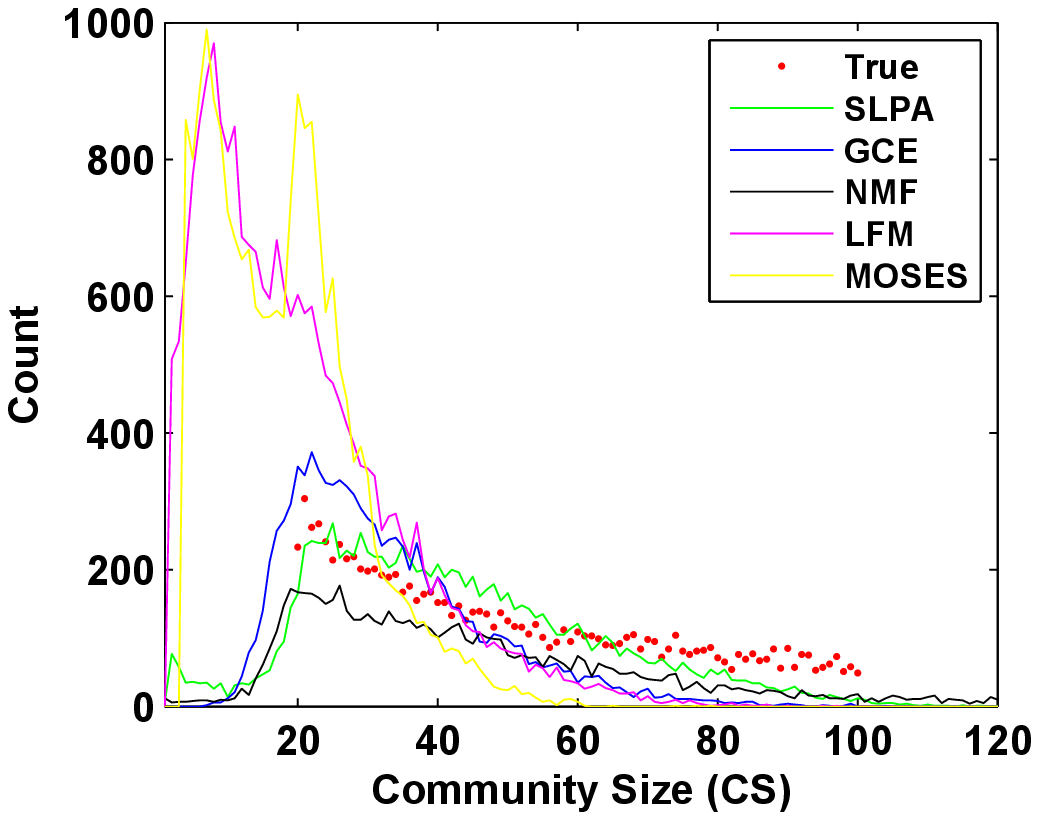} 
		\caption{Histogram of the detected community sizes for SLPA, GCE, NMF, LFM and MOSES created from the results for LFR networks with $n=5000$, $\mu=0.3$ and $O_n=10\%$.}
	 	\label{fig:comdis1}				
	\end{minipage}	
	 \hspace{0.2cm}  
	 \begin{minipage}[t]{0.45\linewidth}
	 \centering
\includegraphics[scale=0.56]{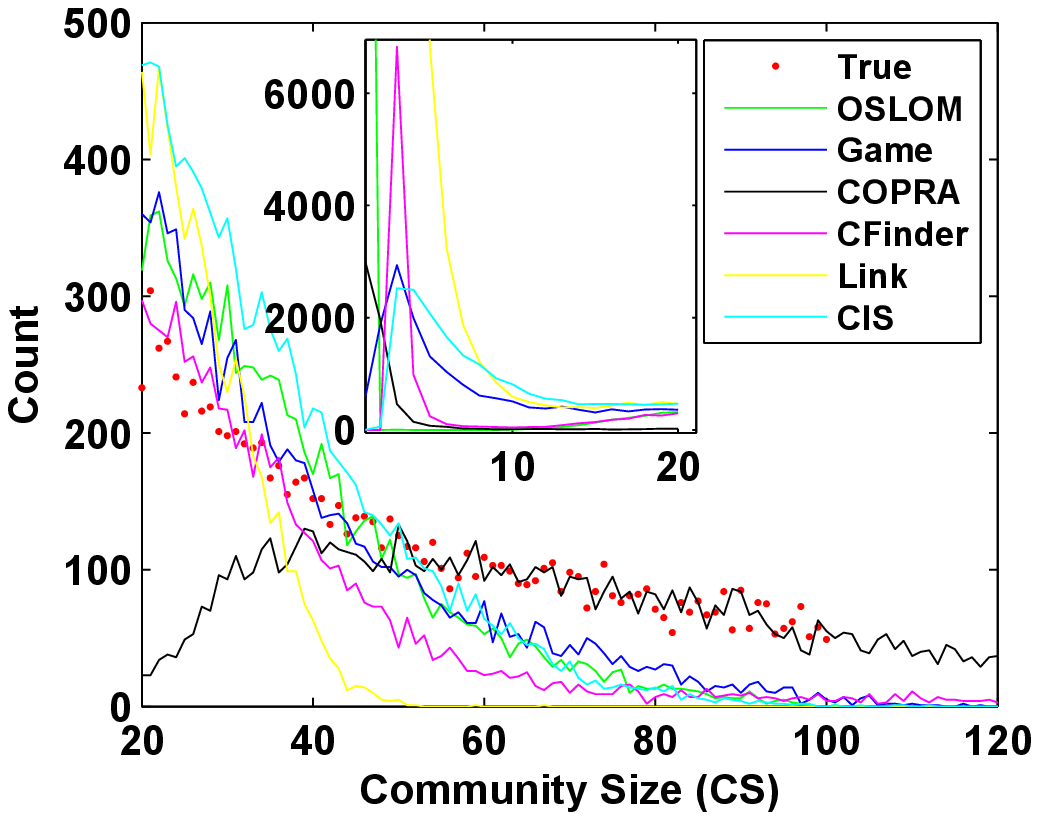} 
		\caption{Histogram of the detected community sizes  for OSLOM, Game, COPRA, CFinder, Link and CIS created from the results for LFR networks with $n=5000$, $\mu=0.3$ and $O_n=10\%$.}
	 	\label{fig:comdis2}	
	\end{minipage}	
	 \hspace{0.2cm}  
	 \begin{minipage}[t]{1.0\linewidth}
	 \vspace{0.4cm}  
	 \centering
	\includegraphics[scale=0.56]{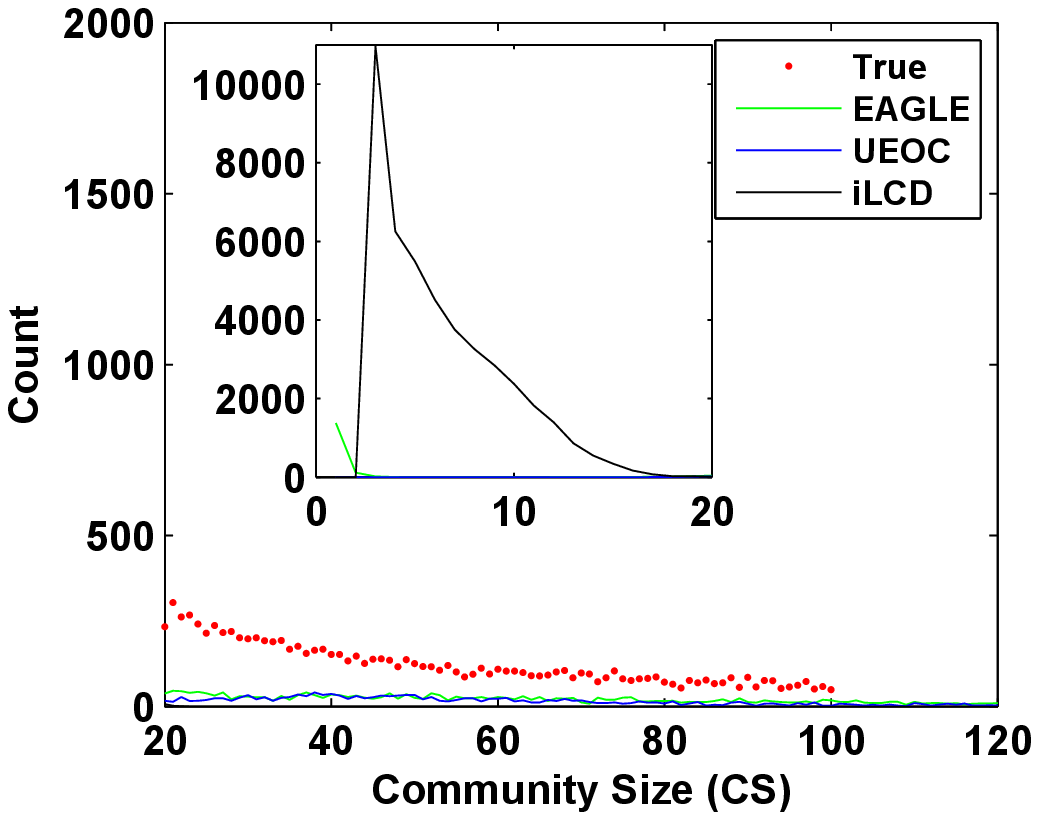} 
		\caption{Histogram of the detected community sizes for EAGLE, UEOC and iLCD crated from the results for LFR networks with $n=5000$, $\mu=0.3$ and $O_n=10\%$.}
	 	\label{fig:comdis3}	
	\end{minipage}	
\end{figure*}
\subsection{Comparing Detected Community Size Distribution in LFR}
\label{sec:comcomdis}
In order to provide insight into the behaviors of different algorithms and verify the ranking, we examined the discovered distribution (histograms) of community sizes (CS) and compared it with the known ground truth. Here we only provide analysis for $n=5000$, $\mu=0.3$, $O_n=10\%$ (the corresponding ranking is $RS^b_{NMI}$ in Table \ref{table:rankingOn01}). In this case, we expect the community size distribution to follow the power law with exponent $\tau_2=1$, a minimum of 20, and a maximum size of 100. Note that the histograms are created from communities over different $O_m's$. As shown in Figure \ref{fig:comdis1}, SLPA, GCE and NMF find communities whose sizes are distributed in a \textit{unimodal} distribution with a single peak at $CS=20$ in aggreement with the ground truth distribution. This explains why they perform well with respect to ranking $RS^b_{NMI}$. LFM and MOSES have a peak around $CS=5$, which lowers their performance. The prominent feature of  Figure \ref{fig:comdis2}  (see the inset) is a \textit{bimodal} distribution that has a peak at $CS=1 \sim 5$. This means that algorithms like OSLOM, Game, COPRA, CIS and CFinder find significant numbers of small communities. In Figure \ref{fig:comdis3}, the distribution is shifted mostly outside the predefined range $20\mathtt{\sim}100$. Algorithms with such a distribution create relatively small communities and perform poorly with respect to this analysis. Here, we conclude that observations on the community size distribution can be used to explain the performance and ranking. 

It is worth noticing that in Figure \ref{fig:comdis2}, excluding the range that contains the undesired peak ($CS =1 \sim 20$), the distributions seem to agree well with the ground truth, especially for COPRA. The performances of OSLOM, Game, CIS and COPRA with respect to NMI are still fairly \textit{stable}. This demonstrates that NMI is, to some degree, not sensitive to small size communities (including outliers or singleton communities). 

\subsection{Identifying Overlapping Nodes in LFR}
\label{sec:identifynodes}
Community overlap manifests itself as the existence of the nodes with membership in multiple communities.  Thus, we will refer to nodes with multiple membership as overlapping nodes. In real-world social networks, such nodes are important because they usually represent bridges (or messengers) between communities. For this reason, the ability to identify overlapping \textit{nodes}, although often \textit{neglected}, is \textit{essential} for assessing the accuracy of community detection algorithms. Measures like NMI and Omega focus only on providing an \textit{overall} measure of algorithmic accuracy. As we see in section \ref{sec:comcomdis}, these measures might not be sensitive enough to provide an accurate picture of what is happening at the \textit{node level}. In this section, we evaluate an algorithm's ability to identify overlapping nodes.

Similar to the definitions of $O_n$ and $O_m$, we define the number of detected overlapping nodes $O_n^d$ and detected memberships $O_m^d$.  Note that the number of overlapping nodes $O_n^d$ alone is insufficient to accurately quantify the detection performance, because it contains both true and false positive. Ideally, an algorithm should report as many true overlapping nodes as possible (i.e., a balance between quality and quantity). To provide more precise analysis, we consider the identification of overlapping nodes as a \textit{binary classification} problem. A node is labeled as \textit{overlapping} as long as $O_m$\textgreater$1$ or $O_m^d$\textgreater$1$ and labeled as \textit{non-overlapping} otherwise. Within this framework, one can use Jaccard index as in \cite{Ball-Jaccard:2011} or F-score as a measure of detection accuracy. In this paper, we use the later that is defined as
\begin{equation}
\label{eq:F}
		F=\frac{2\cdot precision \cdot recall}{precision + recall},
\end{equation}
where \textit{recall} is the number of correctly detected overlapping nodes divided by the true number of overlapping nodes $O_n$, and \textit{precision} is the number of correctly detected overlapping nodes divided by the total number of detected overlapping nodes $O_n^d$. F-score accounts for the balance between detection quantity and quality, and reaches its best and worst value at 1 and 0, respectively. 

\begin{figure}[tp]	
	\centering
    \includegraphics[width=1.0\linewidth]{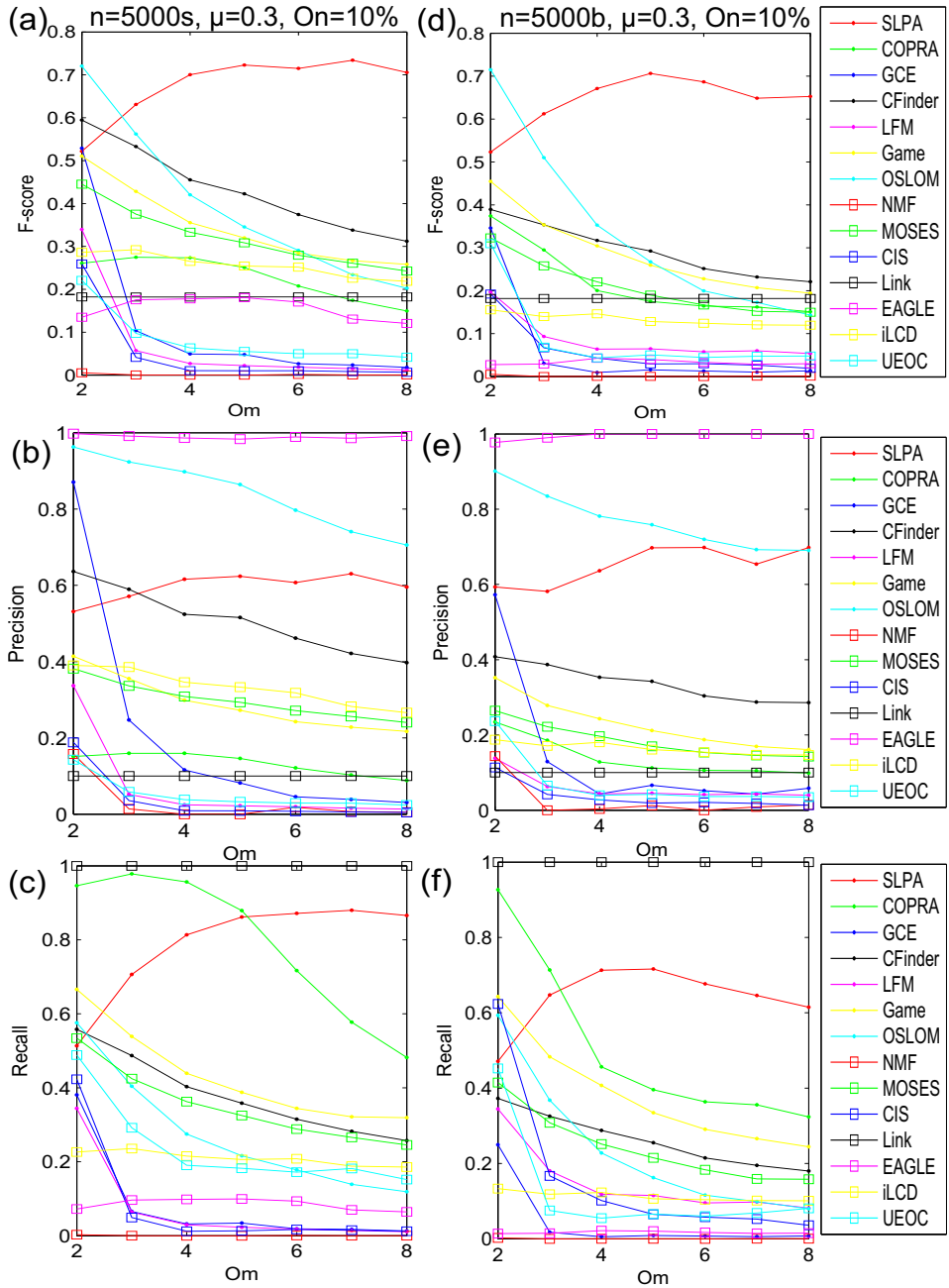} %
		\caption{Evaluations of overlapping node detection on LFR networks with low overlap density $O_n=10\%$. Plots show F-score (together with precision and recall) as a function of the number of memberships for $n=5000$ and $\mu=0.3$. Results for small community size range are shown in the left column, and results for large community size range are shown in the righ column.}
	 	\label{fig:F-NMI-N5000sbMu03On01-all-col}
\end{figure}

\begin{figure}[tp]	
	\centering
    \includegraphics[width=1.0\linewidth]{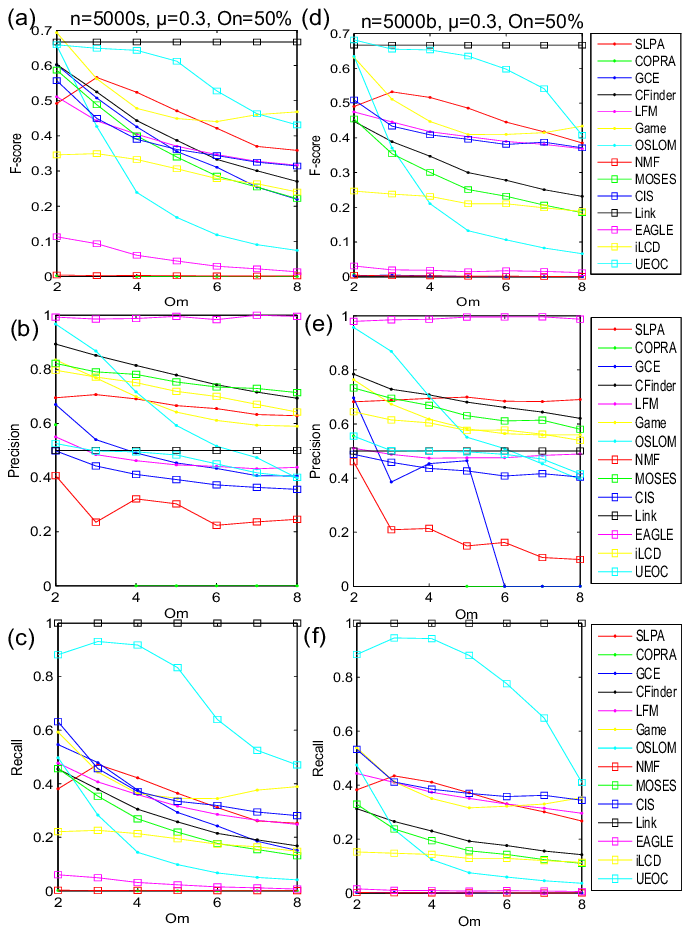} %
		\caption{Evaluations of overlapping node detection on LFR networks with high overlap density $O_n=50\%$. Plots show F-score (together with precision and recall) as a function of the number of memberships for $n=5000$ and $\mu=0.3$. Results for small community size range are shown in the left column, and results for large community size range are shown in the righ column.}
	 	\label{fig:F-NMI-N5000sbMu03On05-all-col}
\end{figure}

Figures \ref{fig:F-NMI-N5000sbMu03On01-all-col} and \ref{fig:F-NMI-N5000sbMu03On05-all-col} show the F-score, precision, and recall for different settings of the LFR benchmark. In general, an algorithm achieves better F-score on benchmarks with small community sizes for both low and high overlapping density. However, the behaviors are quite different for the various algorithms. For example, the gain in F-score on communities in the small size range for EAGLE is due to the increase in recall (i.e., detect more overlapping nodes shown in (c) and (f) in Figure \ref{fig:F-NMI-N5000sbMu03On01-all-col}), while the gain for iLCD is mainly due to the increase in precision (see (b) and (e) in Figure \ref{fig:F-NMI-N5000sbMu03On01-all-col}). Moreover, the F-score (performance) typically decays moderately as overlapping diversity $O_m$ increases. It is evident that $O_m$ has great impacts on OSLOM with an rapid drop for large $O_m$. SLPA is the only exception that has a positive correlation with $O_m$ for low overlapping density case. 

In terms of the precision, half of the algorithms including SLPA, CFinder, Game, OSLOM, MOSES, EAGLE and iLCD consistently outperform the expected random performance, $10\%$ and $50\%$ for low and high overlapping density respectively (see (b) and (e) in both Figures \ref{fig:F-NMI-N5000sbMu03On01-all-col} and \ref{fig:F-NMI-N5000sbMu03On05-all-col}). The high precision of EAGLE (also CFinder and GCE for $O_m=2$) shows that clique-like assumption of communities may help to identify overlapping nodes in low overlapping density case. Link performs merely as well as the random classifier. 
 
\begin{figure}[tp]	
	\centering
      \includegraphics[width=1.0\linewidth]{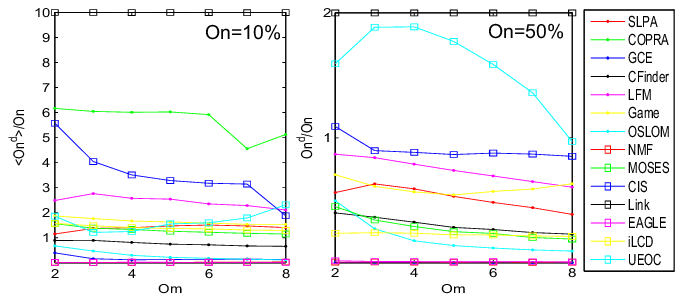} 
		\caption{The number of detected overlapping nodes (normalized by $O_n$) based on the results for LFR networks with $n=5000b$ and $\mu=0.3$. A value larger than 1 is possible. }
	 	\label{fig:numDetOn-all}
\end{figure}

Experiments also reveal an imbalance in precision and recall for some algorithms, which is partially due to either  \textit{over-detection} where more overlapping nodes than there exists are claimed or \textit{under-detection} where only very few overlapping nodes are identified. Extreme examples are EAGLE and Link. Although EAGLE achieves very high detection precision (e.g., (b) and (e) in Figure \ref{fig:F-NMI-N5000sbMu03On01-all-col}), it suffers from under-detection (verified in Figure \ref{fig:numDetOn-all}), which results in a low recall score ((c) and (f) in Figure \ref{fig:F-NMI-N5000sbMu03On01-all-col}).  As a result, we observe a low F-score ((a) and (b) in Figure \ref{fig:F-NMI-N5000sbMu03On01-all-col}). For Link, the algorithm does not perform well in terms of F-score even though it has very high recall ((c) and (f) in Figure \ref{fig:F-NMI-N5000sbMu03On01-all-col}).  This is due to the fact that Link claims way more overlapping nodes than excepted (verified in Figure \ref{fig:numDetOn-all}). 

\subsection{Ranking for Overlapping Node Detection} 
The rankings with respect to F-scores for different community size ranges, $RS^s_{F}$ and $RS^b_{F}$ are shown in Tables \ref{table:node-rankingOn01} and \ref{table:node-rankingOn05} for different overlapping density cases. 
$RS^*_{F}$ is the average ranking over two community size ranges. 

To facilitate comparison, we copy $RS^*_{F}$ into Tables \ref{table:rankingOn01} and \ref{table:rankingOn05}. It is clearly shown, for example in Table \ref{table:rankingOn01}, that the community quality ranking $RS^*_{NMI,Omega}$ and node quality ranking $RS^*_{F}$ might provide quite different pictures of the performance. For the low overlapping density case (as in Table \ref{table:rankingOn01}), algorithms with a low rank in detecting communities could actually have good performances when it comes to identifying overlapping nodes (e.g., CFinder, iLCD and MOSES), while high-ranking algorithms, including GCE and CIS, might perform badly due to under-detection and over-detection respectively. SLPA has very stable and good performance for the low overlapping density case. These observations suggest the need for a careful treatment of the algorithms with a high NMI or Omega score if the application of these algorithms is aimed at identifying nodes with multiple community memberships. Similar conclusions can be drawn for high overlapping density case. 
\begin{table*}[tp]	
\centering
	\begin{minipage}[t]{0.45\linewidth}
	\centering
\centering
\caption{The overlapping node detection ranking for $n=5000$, $\mu=0.3$ and low overlap density $O_n=10\%$.}
\label{table:node-rankingOn01}
\scalebox{0.8} {
	\addtolength{\tabcolsep}{-1pt}  
			\begin{tabular}{c||cc||c} \hline
			\textbf{Rank}	& \textbf{$RS^s_{F}$} & \textbf{$RS^b_{F}$}	& \textbf{$RS^*_{F}$}  \\ \hline
			1&SLPA&SLPA&SLPA \\ 
			2&CFinder&CFinder&CFinder\\ 
			3&OSLOM&Game&Game\\ 
			4&Game&OSLOM&OSLOM\\ 
			5&MOSES&COPRA&MOSES\\ 
			6&iLCD&MOSES&COPRA\\ 
			7&COPRA&Link&iLCD\\ 
			8&Link&iLCD&Link\\ 
			9&EAGLE&LFM&LFM\\ 
			10&GCE&UEOC&UEOC\\ 
			11&UEOC&CIS&EAGLE\\ 
			12&LFM&EAGLE&GCE\\ 
			13&CIS&GCE&CIS\\ 
			14&NMF&NMF&NMF\\ \hline
			\end{tabular}
			}
	\end{minipage}	
	 \hspace{0.2cm}  
	 \begin{minipage}[t]{0.45\linewidth}
	 \centering
\centering
\caption{The overlapping node detection ranking for $n=5000$, $\mu=0.3$ and high overlap density $O_n=50\%$.}
\label{table:node-rankingOn05}
\scalebox{0.8} {
	\addtolength{\tabcolsep}{-1pt}  
			\begin{tabular}{c||cc||c} \hline
			\textbf{Rank}	& \textbf{$RS^s_{F}$} & \textbf{$RS^b_{F}$}	& \textbf{$RS^*_{F}$}  \\ \hline
			1&Link&Link&Link\\ 
			2&UEOC&UEOC&UEOC\\ 
			3&Game&SLPA&SLPA\\ 
			4&SLPA&Game&Game\\ 
			5&CFinder&LFM&LFM\\ 
			6&LFM&CIS&CFinder\\ 
			7&CIS&CFinder&CIS\\ 
			8&GCE&MOSES&MOSES\\ 
			9&MOSES&OSLOM&OSLOM\\ 
			10&iLCD&iLCD&iLCD\\ 
			11&OSLOM&COPRA&GCE\\ 
			12&COPRA&EAGLE&COPRA\\ 
			13&EAGLE&NMF&EAGLE\\ 
			14&NMF&GCE&NMF\\ \hline
			\end{tabular}
			}
	\end{minipage}	
\end{table*}

\subsection{Final Ranking}
Since two types of rankings provide complementary information, we conclude, by considering algorithms that are consistently ranked in the top seven in both $RS^*_{F}$ and $RS^*_{NMI,Omega}$: (a) For low overlapping density networks, SLPA, OSLOM, Game and COPRA offer better performance than the other tested algorithms; (b) For high overlapping density networks, both SLPA and Game provide better performance. (Note that we do not include Link and UEOC because their high ranks are mainly due to the over-detection.)

\section{Tests on Real-world Social Networks}

\begin{figure}[tp]	
	\centering
    \includegraphics*[scale=0.55]{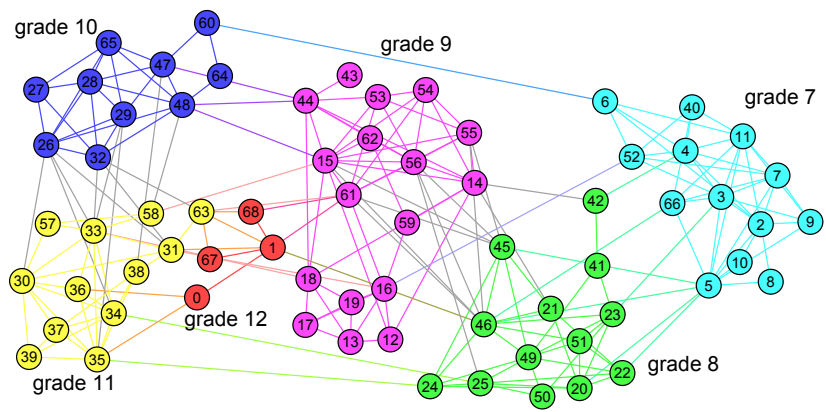} 
		\caption{High school network ($n=69$, $\overline{k}=6.4$). Colors represent known communities corresponding to grades ranging from 7 to 12. Grade 9 is separated into two subgroups that correspond to white (upper) and black (lower) students respectively. Numbers are the node id's. }
	 	\label{fig:comm1}
\end{figure}

We first examined algorithm performance on a high school friendship network\footnote{A project funded by the National Institute of Child Health and Human Development.} where the ground truth is known. This social network from a high school is based on self-reporting from students.  It is known that the true partitioning of the network roughly corresponds to the grade (from 7 to 12) of students listed in the survey. The ground truth is a total of 6 communities (see Figure \ref{fig:comm1}) together with two subgroups within grade 9 corresponding to a group of white and black students. Even though there are no overlapping nodes reported by the students, each algorithm reports some by its own. Results are shown in Table \ref{table:HS}\footnote{For each algorithm, we show results with parameters that output the best NMI score.}. Discovered overlapping nodes are listed in the third column. For algorithms that discover more than 10 overlapping nodes, only the total number is shown. We also include NMI and the number of communities for reference.

It is easy to verify that all the overlapping nodes in Table \ref{table:HS} are connected to at least two different groups. Some of them lie between different grades with strong connections to each individual one, for example, nodes 45, 46, 61, 26, 32 and 33. Some are boundary nodes between subgroups within a grade such as nodes 59, 12 and 18. Node 42 serves as a bridge between groups without having particular coherence to any group. However, it is still not clear whether these nodes are really meaningful or necessary to be considered as ``overlapping". This is one factor that makes the detection (and verification) even more challenging in real-life applications.

\begin{table*}[tp]
\centering
\caption{Test on a high school friendship network.}
\label{table:HS}
\scalebox{0.8} {
	\addtolength{\tabcolsep}{-1pt}  
\begin{tabular}{cccc} \hline
\textbf{Algorithm}	& \textbf{Num. of communities} & \textbf{Overlapping nodes}	& \textbf{NMI}  \\ \hline
CFinder&2&\{12, 18\}&0.1679\\ 
CIS&9&total 34&0.7495\\ 
COPRA&6&total 14&0.7966\\ 
EAGLE&4&\{18\}&0.4962\\ 
Game&10&total 14&0.4673\\ 
GCE&6&\{0, 21, 45, 46, 61\}&0.8333\\ 
iLCD&7&\{5, 21, 26, 29, 31, 32, 33, 46, 61\}&0.3713\\ 
LFM&7&\{0, 45\}&0.8134\\ 
Link&20&total 31&0.3155\\ 
MOSES&10&total 18&0.5037\\ 
NMF&7&\{0, 12, 18, 45\}&0.643\\ 
OSLOM&11&\{45, 46\}&0.4315\\ 
SLPA&6&\{1, 42, 45, 59\}&0.6788\\ 
UEOC&7&\{0, 12, 18, 26, 29, 45\}&0.8148\\ \hline
\end{tabular}
}
\end{table*}

Next, we tested on a wider range of social networks listed in Table \ref{tab:socailNets}. More information about these networks can be found here\footnote{CA-GrQc: a co-authorship network based on papers in General Relativity publishing in Arxiv \cite{dataCA-GrQc}.\\PGP: a network of users of the Pretty-Good-Privacy algorithm \cite{dataPGP}.\\Email: a communication network in Enron via emails \cite{dataSlashdot2008}.\\Epinions: a who-trust-whom on-line social network of a consumer review site Epinions.com \cite{dataEpinion2003}.\\P2P: the Gnutella peer-to-peer file sharing network from August 2002 \cite{dataP2P}.\\Amazon: a co-purchase network of the Amazon website  \cite{dataamazon2003}.\\ Data are available at \url{http://www-personal.umich.edu/~mejn/netdata} and  \url{http://snap.stanford.edu/data}.}. 
Given that the ground truth is not available for most of these networks, we selected two overlapping modularities $Q_{ov}^E$ in (\ref{eq:ShenQov}) and  $Q_{ov}^{Ni}$ in (\ref{eq:NicosiaQov}) as quality measures. The former is based on the node belonging factor, and the later is based on the link belonging factor. For the arbitrary function in $Q_{ov}^{Ni}$, we adopted the one used in \cite{COPRA:2010}, $f(x)=60x-30$.  
\begin{table}[tp]
\centering
\caption{Social networks in the tests}
\label{tab:socailNets}
  \begin{tabular}{cccccc} \hline
			\textbf{Network} & $\mathbf{n}$ &  \textbf{$\overline{k}$} & \textbf{Network} & $\mathbf{n}$ &  \textbf{$\overline{k}$} \\ \hline
			 karate (KR)   	& 34  	& 4.5  & PGP	 	 	 	  	& 10680 & 4.5 \\ 
			 football (FB) 	& 115   & 10.6 & Email (EM)	 		& 33696 & 10.7\\ 
			 lesmis (LS)   	& 77   	& 6.6  & P2P	 	 	 	 		& 62561 & 2.4  \\ 
			 dolphins (DP)	& 62  	& 5.1  & Epinions (EP)		& 75877 & 10.6 \\ 
			 CA-GrQc (CA)		& 4730  & 5.6  & Amazon (AM)	 		& 262111& 6.8 \\ \hline			 
		\end{tabular}
\end{table}

\begin{figure*}[tp]	
\centering
\begin{minipage}[t]{0.48\linewidth}
	 \centering
     \includegraphics[scale=0.44]{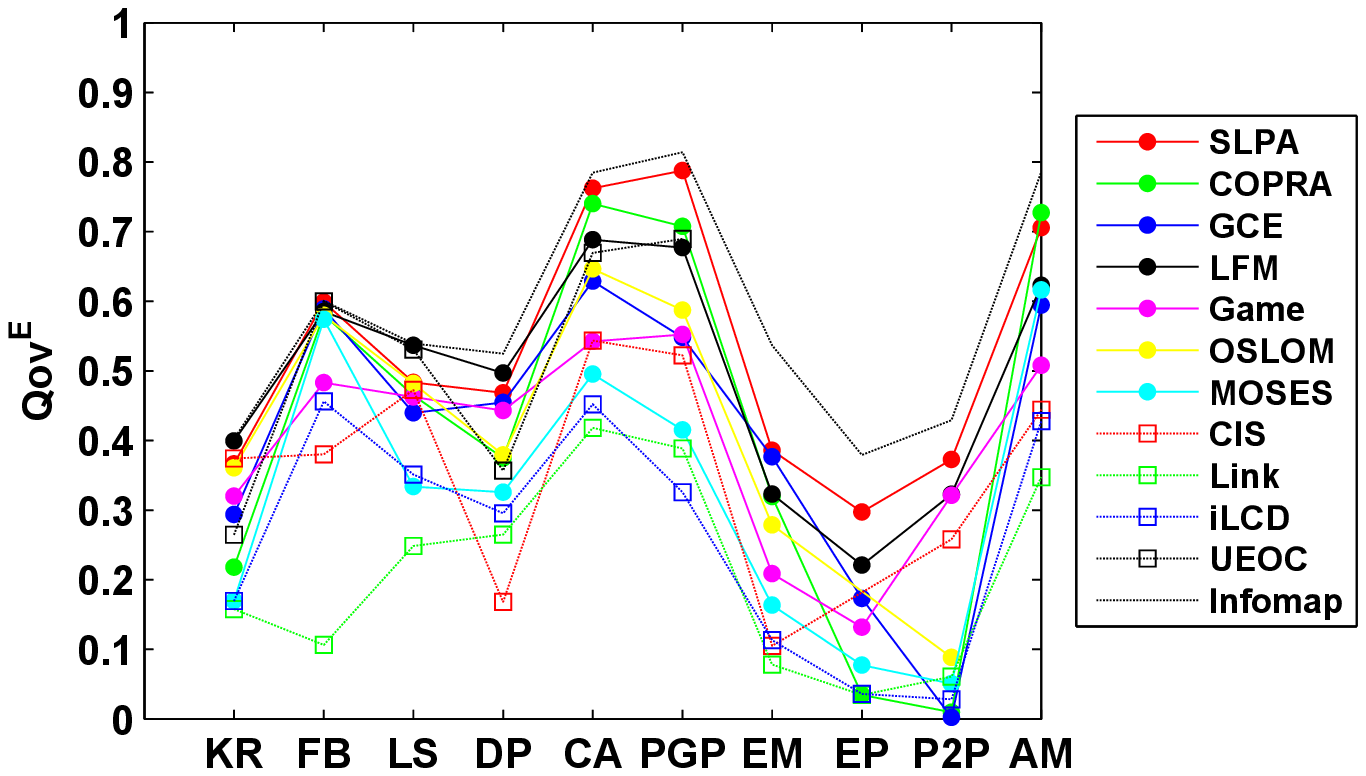} 
		 	\caption{Overlapping modularity $Q_{ov}^E$ for social networks. }
	 	\label{fig:RealQovE}
	\end{minipage}	
	 \hspace{0.1cm}  
	\begin{minipage}[t]{0.48\linewidth}
	\centering
	  \includegraphics[scale=0.44]{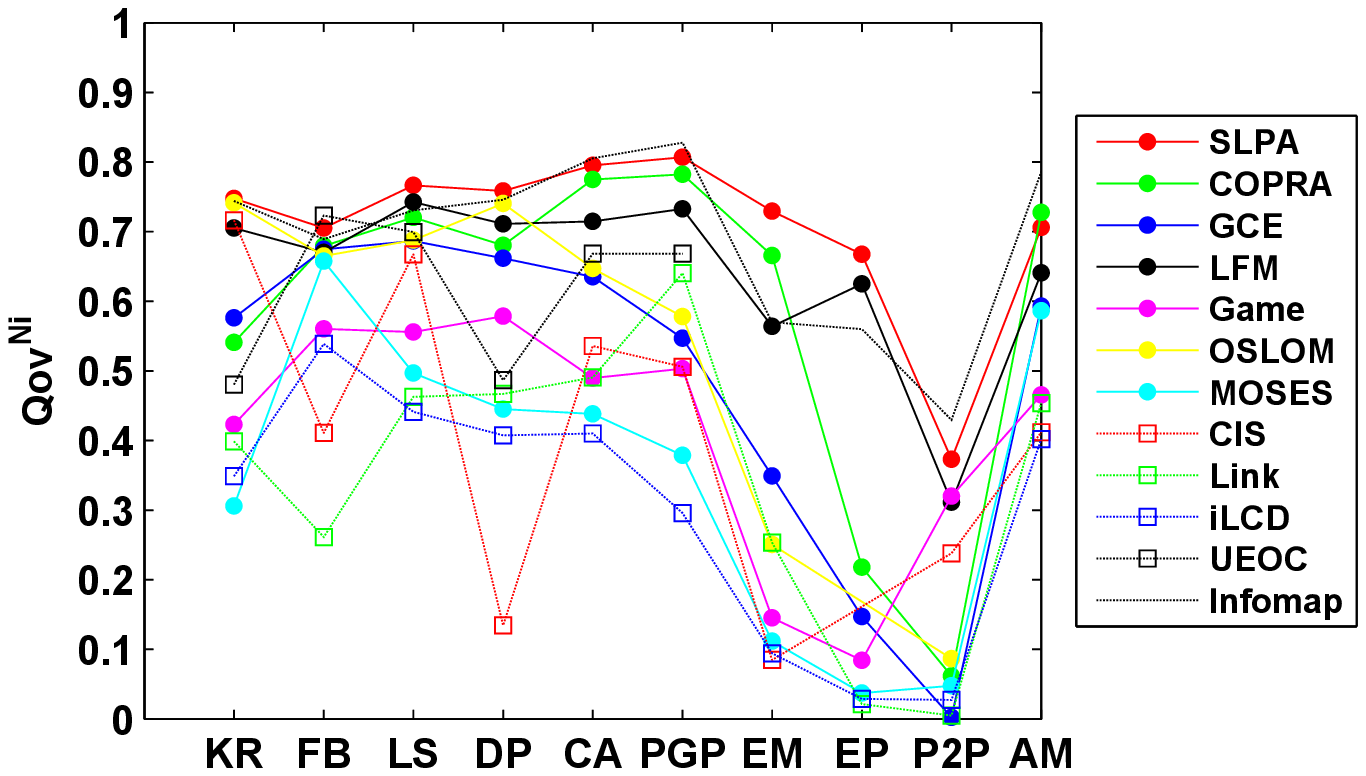} 
		\caption{Overlapping modularity $Q_{ov}^{Ni}$ for social networks.}
	 	\label{fig:RealQovNi}
	\end{minipage}	
\end{figure*}

\begin{figure*}[tp]	
\centering
\begin{minipage}[t]{0.48\linewidth}
	 \centering
     \includegraphics[scale=0.44]{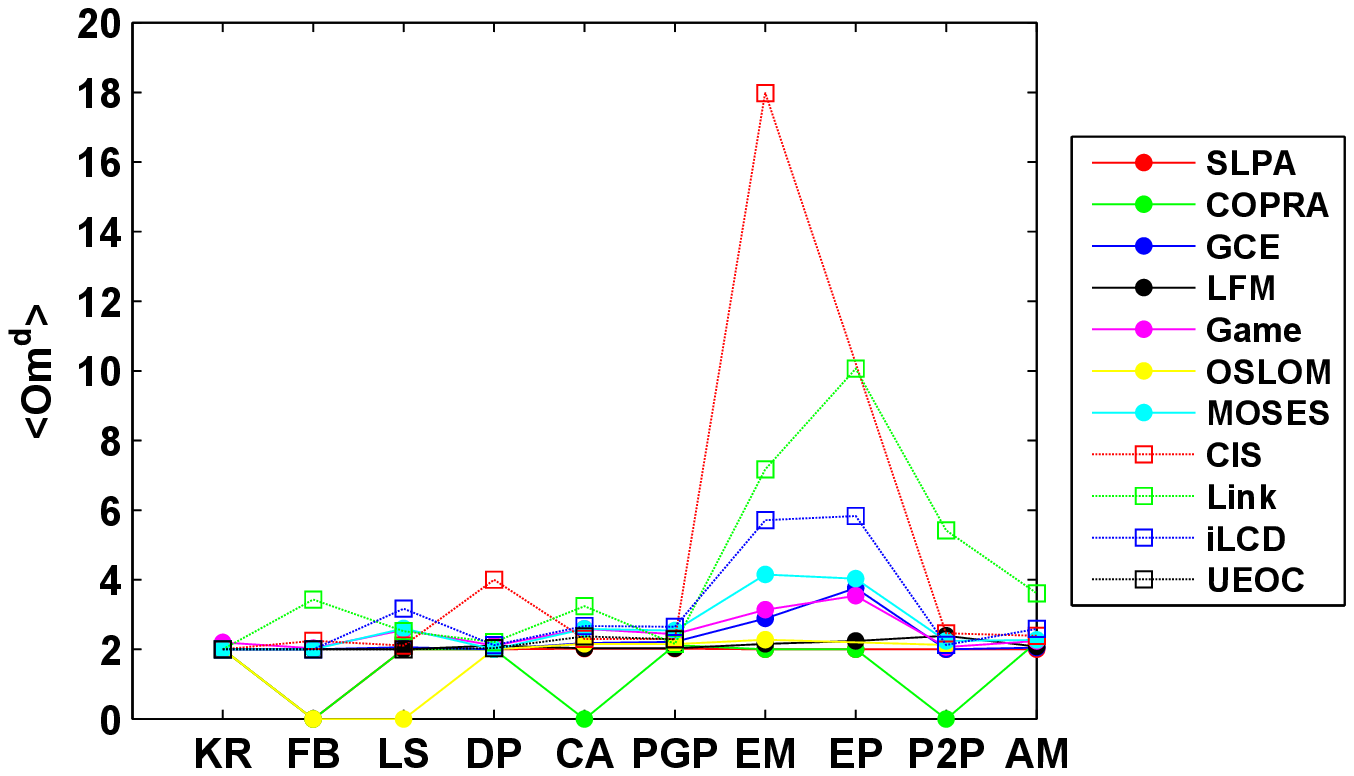} 
		 	\caption{The number of detected memberships for social networks based on the clustering with the best $Q_{ov}^E$. }
	 	\label{fig:RealQovE-Om}
	\end{minipage}	
	 \hspace{0.1cm}  
	\begin{minipage}[t]{0.48\linewidth}
	\centering
	  \includegraphics[scale=0.44]{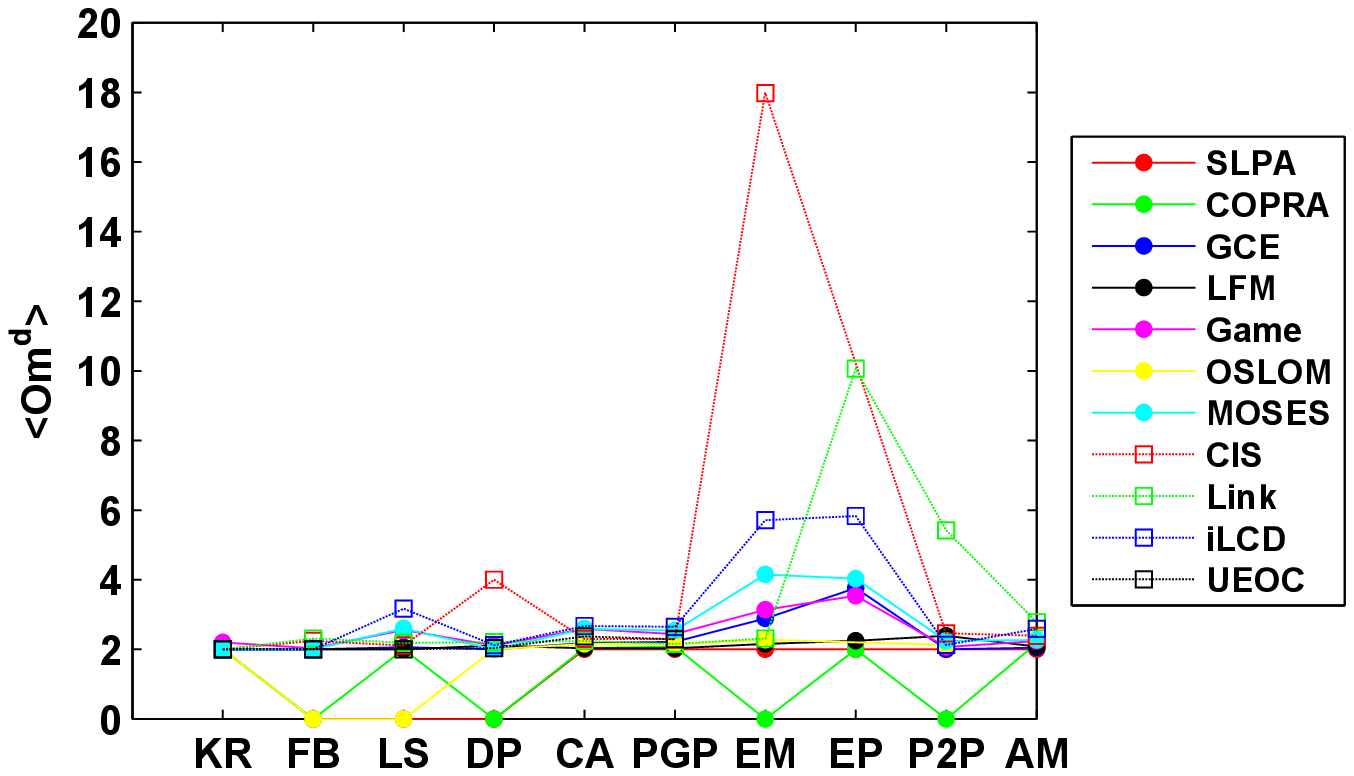} 
		\caption{The number of detected memberships for social networks based on the clustering with  the best $Q_{ov}^{Ni}$. }
	 	\label{fig:RealQovNi-Om}
	\end{minipage}	
\end{figure*}

In Figures \ref{fig:RealQovE} $\sim$ \ref{fig:RealQovNi-Om}, networks are shown in the order of increasing number of edges along the x-axis. Lines connecting points are meant merely to aid the reader in differentiating points from the same algorithm. We removed CFinder, EAGLE and NMF from the test due to either their memory or computation inefficiency in large networks. As a reference, we also performed disjoint community detection with the Infomap  algorithm \cite{Rosvall:2008}, which has been shown to be quite accurate in \cite{LancichinettiComp:2009}.

Figures \ref{fig:RealQovE} and \ref{fig:RealQovNi} show a positive correlation between the two quality measures. Typically, the disjoint partitioning achieves higher $Q_{ov}^E$ than overlapping clusterings, which empirically serves as a bound of the quality of detected overlapping communities. This also holds for $Q_{ov}^{Ni}$ in general. 

\begin{figure*}[tp]	
\centering
\begin{minipage}[t]{0.48\linewidth}
	 \centering
     \includegraphics[scale=0.44]{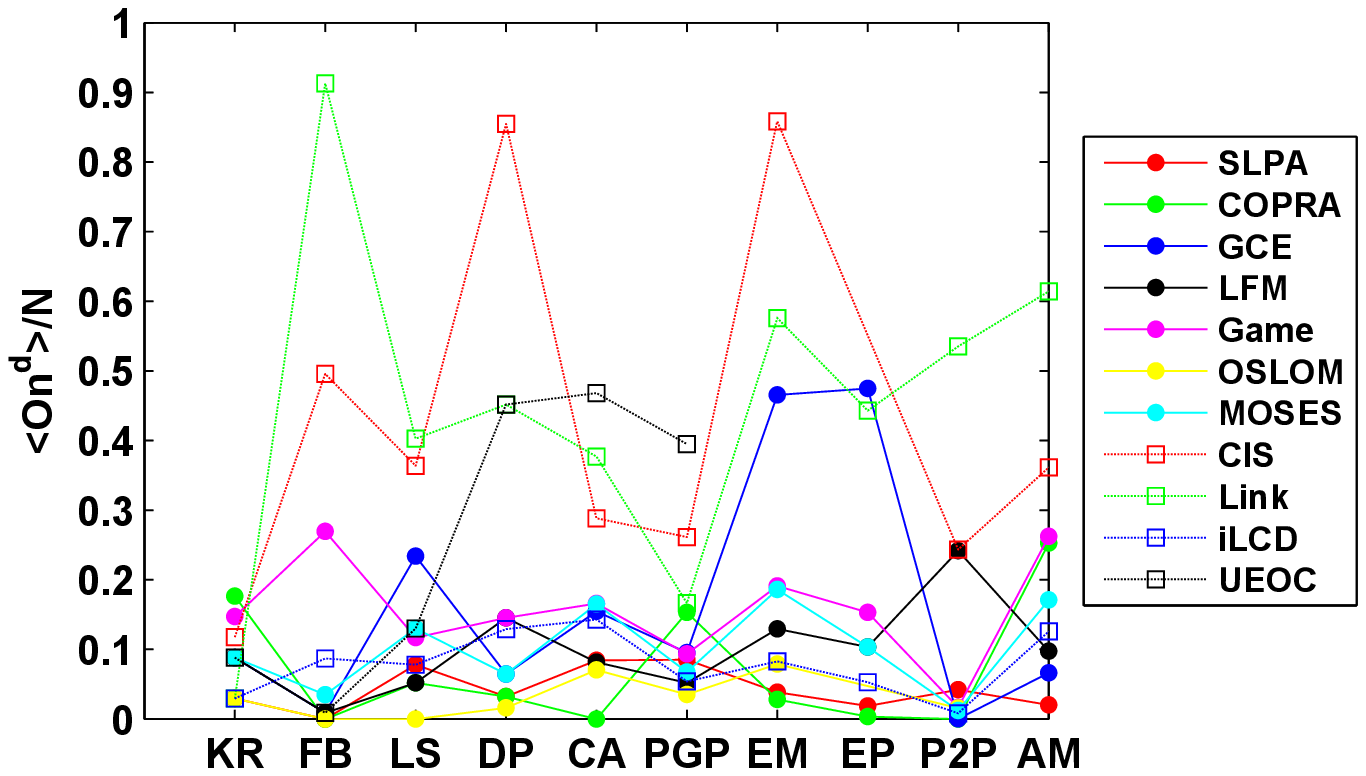} 
		 	\caption{The normalized number of detected overlapping nodes for social networks based on the clustering with the best $Q_{ov}^E$. }
	 	\label{fig:RealQovE-On}
	\end{minipage}	
	 \hspace{0.1cm}  
	\begin{minipage}[t]{0.48\linewidth}
	\centering
	  \includegraphics[scale=0.44]{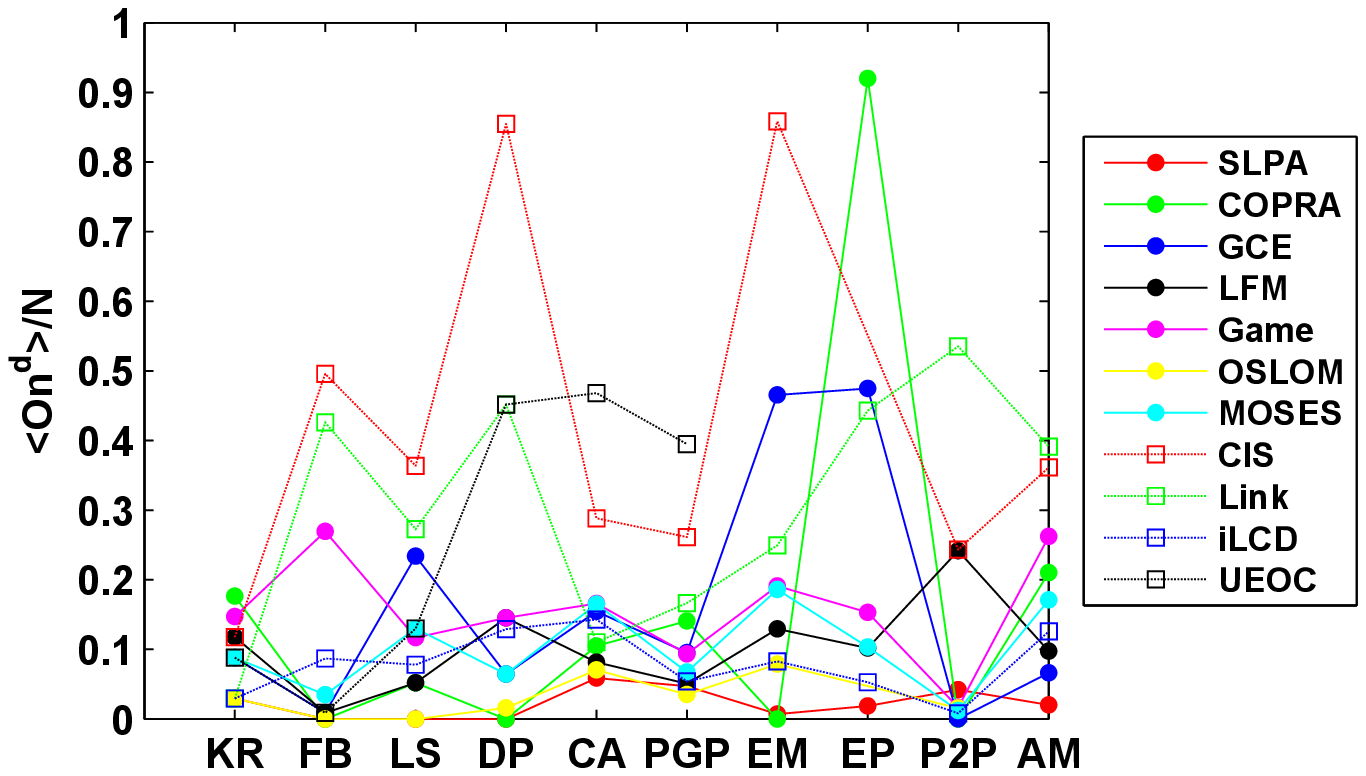} 
		\caption{The normalized number of detected overlapping nodes for social networks based on the clustering with the best $Q_{ov}^{Ni}$. }
	 	\label{fig:RealQovNi-On}
	\end{minipage}	
\end{figure*}

In general, Link and iLCD achieve lower $Q_{ov}^{Ni}$ or $Q_{ov}^E$ compared to others, while SLPA, LFM, COPRA, OSLOM and GCE achieve higher performance on larger networks (e.g., last five networks). Moreover, an algorithms may not perform equally well on different types of network structures. Some of them are sensitive to specific structures. For example, only SLPA, LFM, CIS and Game have satisfying performances in networks with highly sparse structure such as $P2P$, for which COPRA finds merely one single giant community and GCE also fails. Another issue is that some algorithms tend to over-detect the overlap, as was the case for LFR networks. CIS and Link fail in the test because they find too many overlapping nodes or memberships relative to the consensus shown by the other algorithms as seen in Figures \ref{fig:RealQovE-On} $\sim$ \ref{fig:RealQovNi-Om}. Such over-detection happens to other algorithms, including COPRA, GCE and UEOC on specific networks, resulting in low performance for these algorithms.

Some interesting common features are observed from our tests. As shown in Figures \ref{fig:RealQovE-On} and \ref{fig:RealQovNi-On}, the fraction of overlapping nodes found by most of the algorithms is typically less than 30\%. Results from SLPA, OSLOM and COPRA, which offer good performances in the LFR benchmarks, show an even smaller fraction of overlapping nodes, less than 20\%, in most real-world networks examined in this paper. Moreover, Figures \ref{fig:RealQovE-Om} and \ref{fig:RealQovNi-Om} confirm that the diversity (i.e., membership) of overlapping nodes in the tested social networks is relatively small as well, typically 2 or 3 . 


\section{Conclusions and Discussions}
In this paper, we review a wide rang of overlapping community detection algorithms along with quality measures and several existing benchmarks. A number of tests are performed on the LFR benchmarks, incorporating different network structures and various degree of overlapping. Quality evaluation is performed on both community and node levels to provide complementary information. Results show that the detection in networks with high overlapping density and high overlapping diversity has still space for improvements. The node level evaluation reveals the problems of over-detection and under-detection which needs to be considered when designing or evaluating detection algorithms. The results discovered in real-world social networks suggest the sensitivity of some algorithms to sparse networks. A common feature of social networks in view of agreement of different algorithms is relatively small number of overlapping nodes, most of which belong just to a few communities. Moreover, the ambiguity in the definition of overlapping nodes imposes challenges in real-life applications as well.

Here, we review work that has been done mostly for unweighted networks. However, there are number of applications where a weight bears significant information (e.g., the correlation network in biological studies \cite{WGCNA2008}). Algorithms that explicitly take weights into account and allow overlapping, such as CPMw and SLPAw, expect to have advantages over others.

Despite the large amount of work devoted to developing detection algorithms, there are a number of fundamental questions that have yet to be fully addressed. Two of the most prominent are \textit{when to apply overlapping methods} and \textit{how significant the overlapping is}.

It is natural to ask whether or not an application of the overlapping detection algorithms captures any additional information that a disjoint algorithm would necessarily miss. Unfortunately, measures like NMI and Omega do not offer a satisfying answer. The discussion on the necessity of overlap has largely been left unexplored. In \cite{stephenRPI:2009}, the authors empirically examined attributes of the vertices in a network representing commenting activity. The authors suggest that, for a pair of communities $A$ and $B$, the trait similarity between $A\cap B$ and the sets $A-B$ and $B-A$ be higher than the similarity between $A-B$ and $B-A$. Such a relationship might offer a way to estimate the validation of the overlap.

The significance of community structures has been previously explored only within the context of disjoint community detection and based on the notion of modularity \cite{reichardt2006}, \cite{Guimera2004}, \cite{MassenDoye:2005}. The robustness and uniqueness of a discovered partitioning is also examined in \cite{Gfeller2005,Karrer2008,MassenDoye:2007}. Many of these techniques can be extend to assess the overlapping community structure. Interestingly, statistical significance has begun to be included in detection methodologies such as OSLOM \cite{OSLOMLancichinetti:2011}.


\begin{acks}
The work of J.X and B.K.S was supported in part by the Army Research
Laboratory under Cooperative Agreement Number
W911NF-09-2-0053 and by the Office of Naval Research
Grant No. N00014-09-1-0607. The views and
conclusions contained in this document are those of the
authors and should not be interpreted as representing
the official policies either expressed or implied of the
Army Research Laboratory, the Office of Naval Research, or the U.S. Government.

The submitted manuscript has been co-authored by a contractor of the
U.S. Government (S.K.) under contract DE-AC05-00OR22725. Accordingly, the
U.S. Government retains a nonexclusive, royalty-free license to
publish or reproduce the published form of this contribution, or
allows others to do so, for U.S. Government purposes. 

The authors would like to thank researchers who generously provided their software and helped setting up our experiments. We especially thank Steve Gregory for his helpful codes and discussions.

\end{acks}

\bibliographystyle{acmtrans}
\bibliography{bib/CommunityBIB-Jerry-surveyacm}

\begin{thebibliography}{}

\bibitem[\protect\citeauthoryear{Ahn, Bagrow, and Lehmann}{Ahn
  et~al\mbox{.}}{2010}]{YYLinkClustering:2010}
{\sc Ahn, Y.-Y.}, {\sc Bagrow, J.~P.}, {\sc and} {\sc Lehmann, S.} 2010.
\newblock Link communities reveal multiscale complexity in networks.
\newblock {\em Nature\/}~{\em 466}, 761--764.

\bibitem[\protect\citeauthoryear{Ankerst, Breunig, Kriegel, and Sander}{Ankerst
  et~al\mbox{.}}{1999}]{OPTICS:1999}
{\sc Ankerst, M.}, {\sc Breunig, M.~M.}, {\sc Kriegel, H.-P.}, {\sc and} {\sc
  Sander, J.} 1999.
\newblock Optics: ordering points to identify the clustering structure.
\newblock In {\em Proc. SIGKDD Conf.} 49--60.

\bibitem[\protect\citeauthoryear{Arenas, D\'\i{}az-Guilera, and
  P\'erez-Vicente}{Arenas et~al\mbox{.}}{2006}]{AlexSyn:2006}
{\sc Arenas, A.}, {\sc D\'\i{}az-Guilera, A.}, {\sc and} {\sc P\'erez-Vicente,
  C.~J.} 2006.
\newblock Synchronization reveals topological scales in complex networks.
\newblock {\em Phys. Rev. Lett.\/}~{\em 96,\/}~11, 114102.

\bibitem[\protect\citeauthoryear{Ball, Karrer, and Newman}{Ball
  et~al\mbox{.}}{2011}]{Ball-Jaccard:2011}
{\sc Ball, B.}, {\sc Karrer, B.}, {\sc and} {\sc Newman, M. E.~J.} 2011.
\newblock Efficient and principled method for detecting communities in
  networks.
\newblock {\em Phys. Rev. E\/}~{\em 84}, 036103.

\bibitem[\protect\citeauthoryear{Baumes, Goldberg, Krishnamoorthy,
  Magdon-Ismail, and Preston}{Baumes et~al\mbox{.}}{2005}]{Baumes1:2005}
{\sc Baumes, J.}, {\sc Goldberg, M.}, {\sc Krishnamoorthy, M.}, {\sc
  Magdon-Ismail, M.}, {\sc and} {\sc Preston, N.} 2005.
\newblock Finding communities by clustering a graph into overlapping subgraphs.
\newblock In {\em Proc. IADIS}. 97--104.

\bibitem[\protect\citeauthoryear{Bianconi, Pin, and Marsili}{Bianconi
  et~al\mbox{.}}{2008}]{Bianconi:2008}
{\sc Bianconi, G.}, {\sc Pin, P.}, {\sc and} {\sc Marsili, M.} 2008.
\newblock Assessing the relevance of node features for network structure.
\newblock {\em Proc. Natl. Acad. Sci. USA\/}~28, 7.

\bibitem[\protect\citeauthoryear{Blatt, Wiseman, and Domany}{Blatt
  et~al\mbox{.}}{1996}]{Blatt:1996}
{\sc Blatt, M.}, {\sc Wiseman, S.}, {\sc and} {\sc Domany, E.} 1996.
\newblock Superparamagnetic clustering of data.
\newblock {\em Phys. Rev. Lett.\/}~{\em 76}, 3251--3254.

\bibitem[\protect\citeauthoryear{Boguna, Pastor-Satorras, Diaz-Guilera, and
  Arenas}{Boguna et~al\mbox{.}}{2004}]{dataPGP}
{\sc Boguna, M.}, {\sc Pastor-Satorras, R.}, {\sc Diaz-Guilera, A.}, {\sc and}
  {\sc Arenas, A.} 2004.
\newblock Models of social networks based on social distance attachment.
\newblock {\em Phys. Rev. E\/}~{\em 70}, 056122.

\bibitem[\protect\citeauthoryear{Breve, Zhao, and Quiles}{Breve
  et~al\mbox{.}}{2009}]{BreveFabricio:2009}
{\sc Breve, F.}, {\sc Zhao, L.}, {\sc and} {\sc Quiles, M.} 2009.
\newblock Uncovering overlap community structure in complex networks using
  particle competition.
\newblock In {\em Proc. ICAI}. 619--628.

\bibitem[\protect\citeauthoryear{Campello}{Campello}{2007}]{FuzzyExternInd:200%
7}
{\sc Campello, R. J. G.~B.} 2007.
\newblock A fuzzy extension of the rand index and other related indexes for
  clustering and classification assessment.
\newblock {\em Pattern Recogn. Lett.\/}~{\em 28}, 833--841.

\bibitem[\protect\citeauthoryear{Campello.}{Campello.}{2010}]{FuzzyExternInd:2%
010}
{\sc Campello., R. J. G.~B.} 2010.
\newblock Generalized external indexes for comparing data partitions with
  overlapping categories.
\newblock {\em Pattern Recogn. Lett.\/}~{\em 31}, 966--975.

\bibitem[\protect\citeauthoryear{Cazabet, Amblard, and Hanachi}{Cazabet
  et~al\mbox{.}}{2010}]{iLCDCazabet:2010}
{\sc Cazabet, R.}, {\sc Amblard, F.}, {\sc and} {\sc Hanachi, C.} 2010.
\newblock Detection of overlapping communities in dynamical social networks.
\newblock In {\em Proc. SOCIALCOM}. 309--314.

\bibitem[\protect\citeauthoryear{Chen, Shang, Lv, and Fu}{Chen
  et~al\mbox{.}}{2010}]{DuanbingChen:2010}
{\sc Chen, D.}, {\sc Shang, M.}, {\sc Lv, Z.}, {\sc and} {\sc Fu, Y.} 2010.
\newblock Detecting overlapping communities of weighted networks via a local
  algorithm.
\newblock {\em Physica A\/}~{\em 389,\/}~19, 4177--4187.

\bibitem[\protect\citeauthoryear{Chen, Za\"{\i}ane, and Goebel}{Chen
  et~al\mbox{.}}{2009}]{ONDOCS:2009}
{\sc Chen, J.}, {\sc Za\"{\i}ane, O.~R.}, {\sc and} {\sc Goebel, R.} 2009.
\newblock A visual data mining approach to find overlapping communities in
  networks.
\newblock In {\em Proc. ASONAM Conf.} 338--343.

\bibitem[\protect\citeauthoryear{Chen, Liu, Sun, and Wang}{Chen
  et~al\mbox{.}}{2010}]{ChenWei:2010}
{\sc Chen, W.}, {\sc Liu, Z.}, {\sc Sun, X.}, {\sc and} {\sc Wang, Y.} 2010.
\newblock A game-theoretic framework to identify overlapping communities in
  social networks.
\newblock {\em Data Min. Knowl. Discov.\/}~{\em 21}, 224--240.

\bibitem[\protect\citeauthoryear{Collins and Dent}{Collins and
  Dent}{1988}]{Omega:1988}
{\sc Collins, L.~M.} {\sc and} {\sc Dent, C.~W.} 1988.
\newblock Omega: A general formulation of the rand index of cluster recovery
  suitable for non-disjoint solutions.
\newblock {\em Multivar. Behav. Res.\/}~{\em 23,\/}~2 (Feb.), 231--242.

\bibitem[\protect\citeauthoryear{Condon and Karp}{Condon and
  Karp}{2001}]{plantedLpartition:2001}
{\sc Condon, A.} {\sc and} {\sc Karp, R.~M.} 2001.
\newblock Algorithms for graph partitioning on the planted bisection model.
\newblock {\em Rand. Struct. Algo.\/}~{\em 18}, 116--140.

\bibitem[\protect\citeauthoryear{Danon, Duch, Arenas, and Diaz-guilera}{Danon
  et~al\mbox{.}}{2005}]{Danon05comparingcommunity}
{\sc Danon, L.}, {\sc Duch, J.}, {\sc Arenas, A.}, {\sc and} {\sc Diaz-guilera,
  A.} 2005.
\newblock Comparing community structure identification.
\newblock {\em J. Stat. Mech.\/}, 09008.

\bibitem[\protect\citeauthoryear{Davis and Carley}{Davis and
  Carley}{2008}]{DavisGeorge:2008}
{\sc Davis, G.~B.} {\sc and} {\sc Carley, K.} 2008.
\newblock Clearing the fog: Fuzzy, overlapping groups for social networks.
\newblock {\em Social Networks\/}~{\em 30,\/}~3, 201--212.

\bibitem[\protect\citeauthoryear{Ding, Luo, Shi, and Fang}{Ding
  et~al\mbox{.}}{2010}]{AffinityFanDing:2010}
{\sc Ding, F.}, {\sc Luo, Z.}, {\sc Shi, J.}, {\sc and} {\sc Fang, X.} 2010.
\newblock Overlapping community detection by kernel-based fuzzy affinity
  propagation.
\newblock In {\em Proc. ISA Workshop}. 1--4.

\bibitem[\protect\citeauthoryear{Du, Wang, and Wu}{Du
  et~al\mbox{.}}{2008}]{DuNan:2008}
{\sc Du, N.}, {\sc Wang, B.}, {\sc and} {\sc Wu, B.} 2008.
\newblock Overlapping community structure detection in networks.
\newblock In {\em Proc. CIKM}. 1371--1372.

\bibitem[\protect\citeauthoryear{Evans}{Evans}{2010}]{evans:2010}
{\sc Evans, T.} 2010.
\newblock Clique graphs and overlapping communities (arxiv: 1009.0638).

\bibitem[\protect\citeauthoryear{Evans and Lambiotte}{Evans and
  Lambiotte}{2010}]{LineGraph1:2010}
{\sc Evans, T.} {\sc and} {\sc Lambiotte, R.} 2010.
\newblock Line graphs of weighted networks for overlapping communities.
\newblock {\em Eur. Phys. J. B\/}~{\em 77}, 265.

\bibitem[\protect\citeauthoryear{Evans and Lambiotte}{Evans and
  Lambiotte}{2009}]{LineGraph2:2009}
{\sc Evans, T.~S.} {\sc and} {\sc Lambiotte, R.} 2009.
\newblock Line graphs, link partitions and overlapping communities.
\newblock {\em Phys. Rev. E\/}~{\em 80}, 016105.

\bibitem[\protect\citeauthoryear{Farkas, \'Abel, Palla, and Vicsek}{Farkas
  et~al\mbox{.}}{2007}]{CPMwFarkas:2007}
{\sc Farkas, I.}, {\sc \'Abel, D.}, {\sc Palla, G.}, {\sc and} {\sc Vicsek, T.}
  2007.
\newblock Weighted network modules.
\newblock {\em New J. Phys.\/}~{\em 9,\/}~6, 180.

\bibitem[\protect\citeauthoryear{Fisher}{Fisher}{1989}]{Triangle:1989}
{\sc Fisher, D.~C.} 1989.
\newblock Lower bounds on the number of triangles in a graph.
\newblock {\em Journal of Graph Theory\/}~{\em 13,\/}~4, 505--512.

\bibitem[\protect\citeauthoryear{Fortunato}{Fortunato}{2010}]{Santo:2010}
{\sc Fortunato, S.} 2010.
\newblock Community detection in graphs.
\newblock {\em Phys. Rep.\/}~{\em 486}, 75--174.

\bibitem[\protect\citeauthoryear{Frey and Dueck}{Frey and
  Dueck}{2007}]{affinitypropagation:2007}
{\sc Frey, B.~J.} {\sc and} {\sc Dueck, D.} 2007.
\newblock Clustering by passing messages between data points.
\newblock {\em Science\/}~{\em 315}, 972--976.

\bibitem[\protect\citeauthoryear{Fu and Banerjee}{Fu and
  Banerjee}{2008}]{MMMQiangFu:2008}
{\sc Fu, Q.} {\sc and} {\sc Banerjee, A.} 2008.
\newblock Multiplicative mixture models for overlapping clustering.
\newblock In {\em Proc. ICDM}. 791--796.

\bibitem[\protect\citeauthoryear{Geweniger, Z\"{u}hlke, Hammer, and
  Villmann}{Geweniger et~al\mbox{.}}{2009}]{Geweniger:2009}
{\sc Geweniger, T.}, {\sc Z\"{u}hlke, D.}, {\sc Hammer, B.}, {\sc and} {\sc
  Villmann, T.} 2009.
\newblock Fuzzy variant of affinity propagation in comparison to median fuzzy
  c-means.
\newblock In {\em Proc. WSOM}. 72--79.

\bibitem[\protect\citeauthoryear{Gfeller, Chappelier, and De~Los~Rios}{Gfeller
  et~al\mbox{.}}{2005}]{Gfeller2005}
{\sc Gfeller, D.}, {\sc Chappelier, J.-C.}, {\sc and} {\sc De~Los~Rios, P.}
  2005.
\newblock Finding instabilities in the community structure of complex networks.
\newblock {\em Phys. Rev. E\/}~{\em 72}, 056135.

\bibitem[\protect\citeauthoryear{Girvan and Newman}{Girvan and
  Newman}{2002}]{GirNew02}
{\sc Girvan, M.} {\sc and} {\sc Newman, M. E.~J.} 2002.
\newblock Community structure in social and biological networks.
\newblock {\em Proc. Natl. Acad. Sci. USA\/}~{\em 99,\/}~12, 7821--7826.

\bibitem[\protect\citeauthoryear{Gregory}{Gregory}{2007}]{CONGA:2007}
{\sc Gregory, S.} 2007.
\newblock An algorithm to find overlapping community structure in networks.
\newblock In {\em Proc. PKDD Conf.} 91--102.

\bibitem[\protect\citeauthoryear{Gregory}{Gregory}{2008}]{CONGO:2008}
{\sc Gregory, S.} 2008.
\newblock A fast algorithm to find overlapping communities in networks.
\newblock {\em Lect. Notes Comput. Sci.\/}~{\em 5211}, 408.

\bibitem[\protect\citeauthoryear{Gregory}{Gregory}{2009}]{SteveDisjoint:2009}
{\sc Gregory, S.} 2009.
\newblock Finding overlapping communities using disjoint community detection
  algorithms.
\newblock {\em CompleNet\/}~{\em 207}, 47--61.

\bibitem[\protect\citeauthoryear{Gregory}{Gregory}{2010}]{COPRA:2010}
{\sc Gregory, S.} 2010.
\newblock Finding overlapping communities in networks by label propagation.
\newblock {\em New J. Phys.\/}~{\em 12}, 10301.

\bibitem[\protect\citeauthoryear{Gregory}{Gregory}{2011}]{SteveSurvey:2011}
{\sc Gregory, S.} 2011.
\newblock Fuzzy overlapping communities in networks.
\newblock {\em J. Stat. Mech.\/}~{\em 2011,\/}~02, P02017.

\bibitem[\protect\citeauthoryear{Guimer\`a, Sales-Pardo, and Amaral}{Guimer\`a
  et~al\mbox{.}}{2004}]{Guimera2004}
{\sc Guimer\`a, R.}, {\sc Sales-Pardo, M.}, {\sc and} {\sc Amaral, L. A.~N.}
  2004.
\newblock Modularity from fluctuations in random graphs and complex networks.
\newblock {\em Phys. Rev. E\/}~{\em 70}, 025101.

\bibitem[\protect\citeauthoryear{Havemann, Heinz, Struck, and Glaser}{Havemann
  et~al\mbox{.}}{2011}]{MONC:2011}
{\sc Havemann, F.}, {\sc Heinz, M.}, {\sc Struck, A.}, {\sc and} {\sc Glaser,
  J.} 2011.
\newblock Identification of overlapping communities and their hierarchy by
  locally calculating community-changing resolution levels.
\newblock {\em J. Stat. Mech.\/}~{\em 2011,\/}~01, P01023.

\bibitem[\protect\citeauthoryear{Hubert and Arabie}{Hubert and
  Arabie}{1985}]{Hubert:1985}
{\sc Hubert, L.} {\sc and} {\sc Arabie, P.} 1985.
\newblock Comparing partitions.
\newblock {\em Journal of Classification\/}~{\em 2}, 193--218.

\bibitem[\protect\citeauthoryear{H{\"u}llermeier and Rifqi}{H{\"u}llermeier and
  Rifqi}{2009}]{FuzzyRIndex:2009}
{\sc H{\"u}llermeier, E.} {\sc and} {\sc Rifqi, M.} 2009.
\newblock A fuzzy variant of the rand index for comparing clustering
  structures.
\newblock In {\em Proc. IFSA/EUSFLAT Conf.} 1294--1298.

\bibitem[\protect\citeauthoryear{Jin, Yang, Baquero, Liu, He, and Liu}{Jin
  et~al\mbox{.}}{2011}]{UEOCDiJin:2011}
{\sc Jin, D.}, {\sc Yang, B.}, {\sc Baquero, C.}, {\sc Liu, D.}, {\sc He, D.},
  {\sc and} {\sc Liu, J.} 2011.
\newblock A markov random walk under constraint for discovering overlapping
  communities in complex networks.
\newblock {\em J. Stat. Mech.\/}~{\em 2011,\/}~05, P05031.

\bibitem[\protect\citeauthoryear{Karrer, Levina, and Newman}{Karrer
  et~al\mbox{.}}{2008}]{Karrer2008}
{\sc Karrer, B.}, {\sc Levina, E.}, {\sc and} {\sc Newman, M. E.~J.} 2008.
\newblock Robustness of community structure in networks.
\newblock {\em Phys. Rev. E\/}~{\em 77}, 046119.

\bibitem[\protect\citeauthoryear{Kelley}{Kelley}{2009}]{stephenRPI:2009}
{\sc Kelley, S.} 2009.
\newblock The existence and discovery of overlapping communities in large-scale
  networks.
\newblock Ph.D. thesis, Rensselaer Polytechnic Institute, Troy, NY.

\bibitem[\protect\citeauthoryear{Kelley, Goldberg, Magdon-Ismail, Mertsalov,
  and Wallace}{Kelley et~al\mbox{.}}{2011}]{goldbergchapter6:2011}
{\sc Kelley, S.}, {\sc Goldberg, M.}, {\sc Magdon-Ismail, M.}, {\sc Mertsalov,
  K.}, {\sc and} {\sc Wallace, A.} 2011.
\newblock {\em Handbook of Optimization in Complex Networks}.
\newblock Springer, Chapter~6.

\bibitem[\protect\citeauthoryear{Kim and Jeong}{Kim and
  Jeong}{2011}]{Youngdokim:2011}
{\sc Kim, Y.} {\sc and} {\sc Jeong, H.} 2011.
\newblock The map equation for link community (unpublished).

\bibitem[\protect\citeauthoryear{Kov\'acs, Palotai, Szalay, and
  Csermely}{Kov\'acs et~al\mbox{.}}{2010}]{Landscapespalotai:2010}
{\sc Kov\'acs, I.~A.}, {\sc Palotai, R.}, {\sc Szalay, M.}, {\sc and} {\sc
  Csermely, P.} 2010.
\newblock Community landscapes: An integrative approach to determine
  overlapping network module hierarchy, identify key nodes and predict network
  dynamics.
\newblock {\em PLoS ONE\/}~{\em 5,\/}~9, e12528.

\bibitem[\protect\citeauthoryear{Kumpula, Kivel\"a, Kaski, and
  Saram\"aki}{Kumpula et~al\mbox{.}}{2008}]{SCPKumpula:2008}
{\sc Kumpula, J.~M.}, {\sc Kivel\"a, M.}, {\sc Kaski, K.}, {\sc and} {\sc
  Saram\"aki, J.} 2008.
\newblock Sequential algorithm for fast clique percolation.
\newblock {\em Phys. Rev. E\/}~{\em 78,\/}~2, 026109.

\bibitem[\protect\citeauthoryear{Lancichinetti and Fortunato}{Lancichinetti and
  Fortunato}{2009}]{LancichinettiComp:2009}
{\sc Lancichinetti, A.} {\sc and} {\sc Fortunato, S.} 2009.
\newblock Community detection algorithms: a comparative analysis.
\newblock {\em Phys. Rev. E\/}~{\em 80}, 056117.

\bibitem[\protect\citeauthoryear{Lancichinetti, Fortunato, and
  Kert\'esz}{Lancichinetti et~al\mbox{.}}{2009}]{LancichinettiNMI-LFM:2009}
{\sc Lancichinetti, A.}, {\sc Fortunato, S.}, {\sc and} {\sc Kert\'esz, J.}
  2009.
\newblock Detecting the overlapping and hierarchical community structure of
  complex networks.
\newblock {\em New J. Phys.\/}~{\em 11}, 033015.

\bibitem[\protect\citeauthoryear{Lancichinetti, Fortunato, and
  Radicchi}{Lancichinetti et~al\mbox{.}}{2008}]{LFR:2008}
{\sc Lancichinetti, A.}, {\sc Fortunato, S.}, {\sc and} {\sc Radicchi, F.}
  2008.
\newblock Benchmark graphs for testing community detection algorithms.
\newblock {\em Phys. Rev. E\/}~{\em 78}, 046110.

\bibitem[\protect\citeauthoryear{Lancichinetti, Radicchi, Ramasco, and
  Fortunato}{Lancichinetti et~al\mbox{.}}{2011}]{OSLOMLancichinetti:2011}
{\sc Lancichinetti, A.}, {\sc Radicchi, F.}, {\sc Ramasco, J.~J.}, {\sc and}
  {\sc Fortunato, S.} 2011.
\newblock Finding statistically significant communities in networks.
\newblock {\em PLoS ONE\/}~{\em 6,\/}~4, e18961.

\bibitem[\protect\citeauthoryear{Langfelder and Horvath}{Langfelder and
  Horvath}{2008}]{WGCNA2008}
{\sc Langfelder, P.} {\sc and} {\sc Horvath, S.} 2008.
\newblock {WGCNA}: an {R} package for weighted correlation network analysis.
\newblock {\em BMC Bioinformatics\/}~1, 559.

\bibitem[\protect\citeauthoryear{Latouche, Birmele, and Ambroise}{Latouche
  et~al\mbox{.}}{2011}]{OSBMPierre:2011}
{\sc Latouche, P.}, {\sc Birmele, E.}, {\sc and} {\sc Ambroise, C.} 2011.
\newblock Overlapping stochastic block models with application to the french
  political blogosphere.
\newblock {\em The Annals of Applied Statistics\/}~{\em 5}, 309--336.

\bibitem[\protect\citeauthoryear{Lee, Reid, McDaid, and Hurley}{Lee
  et~al\mbox{.}}{2010}]{GCE:2010}
{\sc Lee, C.}, {\sc Reid, F.}, {\sc McDaid, A.}, {\sc and} {\sc Hurley, N.}
  2010.
\newblock Detecting highly overlapping community structure by greedy clique
  expansion.
\newblock In {\em Proc. SNAKDD Workshop}. 33--42.

\bibitem[\protect\citeauthoryear{Leskovec, Adamic, and Huberman}{Leskovec
  et~al\mbox{.}}{2007}]{dataamazon2003}
{\sc Leskovec, J.}, {\sc Adamic, L.~A.}, {\sc and} {\sc Huberman, B.~A.} 2007.
\newblock The dynamics of viral marketing.
\newblock {\em ACM Trans. Web\/}~{\em 1}, 5.

\bibitem[\protect\citeauthoryear{Leskovec, Kleinberg, and Faloutsos}{Leskovec
  et~al\mbox{.}}{2007}]{dataCA-GrQc}
{\sc Leskovec, J.}, {\sc Kleinberg, J.}, {\sc and} {\sc Faloutsos, C.} 2007.
\newblock Graph evolution: Densification and shrinking diameters.
\newblock {\em ACM TKDD\/}~{\em 1}, 2.

\bibitem[\protect\citeauthoryear{Leskovec, Lang, and andMichael
  W.~Mahoney}{Leskovec et~al\mbox{.}}{2009}]{dataSlashdot2008}
{\sc Leskovec, J.}, {\sc Lang, K.~J.}, {\sc and} {\sc andMichael W.~Mahoney,
  A.~D.} 2009.
\newblock Community structure in large networks: Natural cluster sizes and the
  absence of large well-defined clusters.
\newblock {\em Internet Mathematics\/}~{\em 6}, 29--123.

\bibitem[\protect\citeauthoryear{Leskovec, Lang, and Mahoney}{Leskovec
  et~al\mbox{.}}{2010}]{wwwLeskovec:2010}
{\sc Leskovec, J.}, {\sc Lang, K.~J.}, {\sc and} {\sc Mahoney, M.} 2010.
\newblock Empirical comparison of algorithms for network community detection.
\newblock In {\em Proc. WWW Conf.} 631--640.

\bibitem[\protect\citeauthoryear{Li, Leyva, Almendral, Sendina-Nadal, Buldu,
  Havlin, and Boccaletti}{Li et~al\mbox{.}}{2008}]{synDataLi:2008}
{\sc Li, D.}, {\sc Leyva, I.}, {\sc Almendral, J.}, {\sc Sendina-Nadal, I.},
  {\sc Buldu, J.}, {\sc Havlin, S.}, {\sc and} {\sc Boccaletti, S.} 2008.
\newblock Synchronization interfaces and overlapping communities in complex
  networks.
\newblock {\em Phys. Rev. Lett.\/}~{\em 101}, 168701.

\bibitem[\protect\citeauthoryear{Lu, Korniss, and Szymanski}{Lu
  et~al\mbox{.}}{2009}]{Lu:2009}
{\sc Lu, Q.}, {\sc Korniss, G.}, {\sc and} {\sc Szymanski, B.~K.} 2009.
\newblock The naming game in social networks: community formation and consensus
  engineering.
\newblock {\em J. Econ. Interact. Coord.\/}~{\em 4}, 221--235.

\bibitem[\protect\citeauthoryear{M.~Rosvall}{M.~Rosvall}{2008}]{Rosvall:2008}
{\sc M.~Rosvall, C.~B.} 2008.
\newblock Maps of random walks on complex networks reveal community structure.
\newblock {\em Proc. Natl. Acad. Sci.\/}~{\em 105}, 1118--1123.

\bibitem[\protect\citeauthoryear{Magdon-ismail and Purnell}{Magdon-ismail and
  Purnell}{2011}]{SSDEJonathan:2011}
{\sc Magdon-ismail, M.} {\sc and} {\sc Purnell, J.} 2011.
\newblock Fast overlapping clustering of networks using sampled spectral
  distance embedding and gmms.
\newblock Tech. rep., Rensselaer Polytechnic Institute.

\bibitem[\protect\citeauthoryear{Massen and Doye}{Massen and
  Doye}{2005}]{MassenDoye:2005}
{\sc Massen, C.} {\sc and} {\sc Doye, J.} 2005.
\newblock Identifying communities within energy landscapes.
\newblock {\em Phys. Rev. E\/}~{\em 71}, 046101.

\bibitem[\protect\citeauthoryear{Massen and Doye}{Massen and
  Doye}{2007}]{MassenDoye:2007}
{\sc Massen, C.} {\sc and} {\sc Doye, J.} 2007.
\newblock Thermodynamics of community structure.
\newblock {\em Preprint arXiv:cond-mat/0610077v1\/}.

\bibitem[\protect\citeauthoryear{McDaid and Hurley}{McDaid and
  Hurley}{2010}]{MOSES:2010}
{\sc McDaid, A.} {\sc and} {\sc Hurley, N.} 2010.
\newblock Detecting highly overlapping communities with model-based overlapping
  seed expansion.
\newblock In {\em Proc. ASONAM Conf.} 112--119.

\bibitem[\protect\citeauthoryear{Molloy and Reed}{Molloy and
  Reed}{1995}]{Molloy:1995}
{\sc Molloy, M.} {\sc and} {\sc Reed, B.} 1995.
\newblock A critical point for random graphs with a given degree sequence.
\newblock {\em Rand. Struct. Algo.\/}~{\em 6}, 161--179.

\bibitem[\protect\citeauthoryear{Moon and Moser}{Moon and
  Moser}{1965}]{Cliques:1965}
{\sc Moon, J.} {\sc and} {\sc Moser, L.} 1965.
\newblock On cliques in graphs.
\newblock {\em Israel Journal of Mathematics\/}~{\em 3}, 23--28.

\bibitem[\protect\citeauthoryear{Nepusz, Petr\'oczi, N\'egyessy, and
  Bazs\'o}{Nepusz et~al\mbox{.}}{2008}]{Nepusz:2008}
{\sc Nepusz, T.}, {\sc Petr\'oczi, A.}, {\sc N\'egyessy, L.}, {\sc and} {\sc
  Bazs\'o, F.} 2008.
\newblock Fuzzy communities and the concept of bridgeness in complex networks.
\newblock {\em Phys. Rev. E\/}~{\em 77}, 016107.

\bibitem[\protect\citeauthoryear{Newman}{Newman}{2006}]{PhysRevE.74.036104}
{\sc Newman, M. E.~J.} 2006.
\newblock Finding community structure in networks using the eigenvectors of
  matrices.
\newblock {\em Phys. Rev. E\/}~{\em 74}, 036104.

\bibitem[\protect\citeauthoryear{Newman and Leicht}{Newman and
  Leicht}{2007}]{MixtureNewMan:2007}
{\sc Newman, M. E.~J.} {\sc and} {\sc Leicht, E.~A.} 2007.
\newblock Mixture models and exploratory analysis in networks.
\newblock {\em Proc. Natl. Acad. Sci. USA\/}~{\em 104}, 9564--9569.

\bibitem[\protect\citeauthoryear{Newman, Strogatz, and Watts}{Newman
  et~al\mbox{.}}{2001}]{ANNNewman:2001}
{\sc Newman, M. E.~J.}, {\sc Strogatz, S.~H.}, {\sc and} {\sc Watts, D.~J.}
  2001.
\newblock Random graphs with arbitrary degree distributions and their
  applications.
\newblock {\em Phys. Rev. E\/}~{\em 64,\/}~2, 026118.

\bibitem[\protect\citeauthoryear{Nicosia, Mangioni, Carchiolo, and
  Malgeri}{Nicosia et~al\mbox{.}}{2009}]{Nicosia:2009}
{\sc Nicosia, V.}, {\sc Mangioni, G.}, {\sc Carchiolo, V.}, {\sc and} {\sc
  Malgeri, M.} 2009.
\newblock Extending the definition of modularity to directed graphs with
  overlapping communities.
\newblock {\em J. Stat. Mech.\/}, 03024.

\bibitem[\protect\citeauthoryear{Nowicki and Snijders}{Nowicki and
  Snijders}{2001}]{SBMNowicki:2001}
{\sc Nowicki, K.} {\sc and} {\sc Snijders, T. A.~B.} 2001.
\newblock Estimation and prediction for stochastic blockstructures.
\newblock {\em JASA\/}~{\em 96,\/}~455, 1077--1087.

\bibitem[\protect\citeauthoryear{Padrol-Sureda, Perarnau-Llobet, Pfeifle, and
  Munt¨¦s-Mulero}{Padrol-Sureda et~al\mbox{.}}{2010}]{OCApadrol-sureda:2010}
{\sc Padrol-Sureda, A.}, {\sc Perarnau-Llobet, G.}, {\sc Pfeifle, J.}, {\sc
  and} {\sc Munt¨¦s-Mulero, V.} 2010.
\newblock Overlapping community search for social networks.
\newblock In {\em Proc. ICDE}. 992--995.

\bibitem[\protect\citeauthoryear{Palla, Der\'enyi, Farkas, and Vicsek}{Palla
  et~al\mbox{.}}{2005}]{CPM:2005}
{\sc Palla, G.}, {\sc Der\'enyi, I.}, {\sc Farkas, I.}, {\sc and} {\sc Vicsek,
  T.} 2005.
\newblock Uncovering the overlapping community structure of complex networks in
  nature and society.
\newblock {\em Nature\/}~{\em 435}, 814--818.

\bibitem[\protect\citeauthoryear{Psorakis, Roberts, Ebden, and
  Sheldon}{Psorakis et~al\mbox{.}}{2011}]{BayesianNMFIoannis:2011}
{\sc Psorakis, I.}, {\sc Roberts, S.}, {\sc Ebden, M.}, {\sc and} {\sc Sheldon,
  B.} 2011.
\newblock Overlapping community detection using bayesian non-negative matrix
  factorization.
\newblock {\em Phys. Rev. E\/}~{\em 83,\/}~6, 066114.

\bibitem[\protect\citeauthoryear{Raghavan, Albert, and Kumara}{Raghavan
  et~al\mbox{.}}{2007}]{Raghavan:2007}
{\sc Raghavan, U.~N.}, {\sc Albert, R.}, {\sc and} {\sc Kumara, S.} 2007.
\newblock Near linear time algorithm to detect community structures in
  large-scale networks.
\newblock {\em Phys. Rev. E\/}~{\em 76}, 036106.

\bibitem[\protect\citeauthoryear{Rees and Gallagher}{Rees and
  Gallagher}{2010}]{Rees:2010}
{\sc Rees, B.} {\sc and} {\sc Gallagher, K.} 2010.
\newblock Overlapping community detection by collective friendship group
  inference.
\newblock In {\em Proc. ASONAM Conf.} 375--379.

\bibitem[\protect\citeauthoryear{Reichardt and Bornholdt}{Reichardt and
  Bornholdt}{2004}]{Reichardt:2004}
{\sc Reichardt, J.} {\sc and} {\sc Bornholdt, S.} 2004.
\newblock Detecting fuzzy community structures in complex networks with a potts
  model.
\newblock {\em Phys. Rev. Lett.\/}~{\em 93}, 218701.

\bibitem[\protect\citeauthoryear{Reichardt. and Bornholdt}{Reichardt. and
  Bornholdt}{2006}]{Reichardt:2006}
{\sc Reichardt., J.} {\sc and} {\sc Bornholdt, S.} 2006.
\newblock Statistical mechanics of community detection.
\newblock {\em Phys. Rev. E\/}~{\em 74,\/}~1, 016110.

\bibitem[\protect\citeauthoryear{Reichardt and Bornholdt}{Reichardt and
  Bornholdt}{2006}]{reichardt2006}
{\sc Reichardt, J.} {\sc and} {\sc Bornholdt, S.} 2006.
\newblock {When are networks truly modular?}
\newblock {\em Physica D\/}~{\em 224}, 20--26.

\bibitem[\protect\citeauthoryear{Reid, McDaid, and Hurley}{Reid
  et~al\mbox{.}}{2011}]{Fergal:2011}
{\sc Reid, F.}, {\sc McDaid, A.~F.}, {\sc and} {\sc Hurley, N.~J.} 2011.
\newblock Partitioning breaks communities.
\newblock In {\em Proc. ASONAM Conf.} 102--109.

\bibitem[\protect\citeauthoryear{Ren, Yan, Liao, and Xiao}{Ren
  et~al\mbox{.}}{2009}]{SPAEMRenWei:2009}
{\sc Ren, W.}, {\sc Yan, G.}, {\sc Liao, X.}, {\sc and} {\sc Xiao, L.} 2009.
\newblock Simple probabilistic algorithm for detecting community structure.
\newblock {\em Phys. Rev. E\/}~{\em 79,\/}~3, 036111.

\bibitem[\protect\citeauthoryear{Richardson, Agrawal, and Domingos}{Richardson
  et~al\mbox{.}}{2003}]{dataEpinion2003}
{\sc Richardson, M.}, {\sc Agrawal, R.}, {\sc and} {\sc Domingos, P.} 2003.
\newblock Trust management for the semantic web.
\newblock In {\em Proc. ISWC}. Vol. 2870. 351--368.

\bibitem[\protect\citeauthoryear{Ripeanu, Foster, and Iamnitchi}{Ripeanu
  et~al\mbox{.}}{2002}]{dataP2P}
{\sc Ripeanu, M.}, {\sc Foster, I.}, {\sc and} {\sc Iamnitchi, A.} 2002.
\newblock Mapping the gnutella network: Properties of large-scale peer-to-peer
  systems and implications for system design.
\newblock {\em IEEE Internet Computing Journal\/}~{\em 6}, 1.

\bibitem[\protect\citeauthoryear{Ronhovde and Nussinov}{Ronhovde and
  Nussinov}{2009}]{Peter:2009}
{\sc Ronhovde, P.} {\sc and} {\sc Nussinov, Z.} 2009.
\newblock Multiresolution community detection for megascale networks by
  information-based replica correlations.
\newblock {\em Phys. Rev. E\/}~{\em 80}, 016109.

\bibitem[\protect\citeauthoryear{Sawardecker, Sales-Pardo, and
  Amaral}{Sawardecker et~al\mbox{.}}{2009}]{Sawardecker:2009}
{\sc Sawardecker, E.}, {\sc Sales-Pardo, M.}, {\sc and} {\sc Amaral, L.} 2009.
\newblock Detection of node group membership in networks with group overlap.
\newblock {\em Eur. Phys. J. B\/}~{\em 67}, 277.

\bibitem[\protect\citeauthoryear{Shen, Cheng, Cai, and Hu}{Shen
  et~al\mbox{.}}{2009}]{EAGLE:2009}
{\sc Shen, H.}, {\sc Cheng, X.}, {\sc Cai, K.}, {\sc and} {\sc Hu, M.-B.} 2009.
\newblock Detect overlapping and hierarchical community structure.
\newblock {\em Physica A\/}~{\em 388}, 1706.

\bibitem[\protect\citeauthoryear{Shen, Cheng, and Guo}{Shen
  et~al\mbox{.}}{2009}]{EAGLE:2009b}
{\sc Shen, H.}, {\sc Cheng, X.}, {\sc and} {\sc Guo, J.} 2009.
\newblock Quantifying and identifying the overlapping community structure in
  networks.
\newblock {\em J. Stat. Mech.\/}~07, 9.

\bibitem[\protect\citeauthoryear{Wang, Jiao, and Wu}{Wang
  et~al\mbox{.}}{2009}]{WangLineGraph3:2009}
{\sc Wang, X.}, {\sc Jiao, L.}, {\sc and} {\sc Wu, J.} 2009.
\newblock Adjusting from disjoint to overlapping community detection of complex
  networks.
\newblock {\em Physica A\/}~{\em 388}, 5045--5056.

\bibitem[\protect\citeauthoryear{White and Smyth}{White and
  Smyth}{2005}]{WhiteSmyth:2005}
{\sc White, S.} {\sc and} {\sc Smyth, P.} 2005.
\newblock A spectral clustering approach to finding communities in graphs.
\newblock In {\em Proc. SIAM International Conference on Data Mining}. 76--84.

\bibitem[\protect\citeauthoryear{Wu, Lin, Wan, and Tian}{Wu
  et~al\mbox{.}}{2010}]{Zhihao:2010}
{\sc Wu, Z.}, {\sc Lin, Y.}, {\sc Wan, H.}, {\sc and} {\sc Tian, S.} 2010.
\newblock A fast and reasonable method for community detection with adjustable
  extent of overlapping.
\newblock In {\em Proc. ISKE Conf.} 376--379.

\bibitem[\protect\citeauthoryear{Xie and Szymanski}{Xie and
  Szymanski}{2011}]{JieruiXieLPA:2010}
{\sc Xie, J.} {\sc and} {\sc Szymanski, B.~K.} 2011.
\newblock Community detection using a neighborhood strength driven label
  propagation algorithm.
\newblock In {\em Proc. NSW}. 188--195.

\bibitem[\protect\citeauthoryear{Xie and Szymanski}{Xie and
  Szymanski}{2012}]{JieruiXieSLPA-pkdd:2012}
{\sc Xie, J.} {\sc and} {\sc Szymanski, B.~K.} 2012.
\newblock Towards linear time overlapping community detection in social
  networks.
\newblock In {\em Proc. PAKDD Conf.} 25--36.

\bibitem[\protect\citeauthoryear{Xie, Szymanski, and Liu}{Xie
  et~al\mbox{.}}{2011}]{JieruiXieSLPA-ICDM:2011}
{\sc Xie, J.}, {\sc Szymanski, B.~K.}, {\sc and} {\sc Liu, X.} 2011.
\newblock {SLPA}: Uncovering overlapping communities in social networks via a
  speaker-listener interaction dynamic process.
\newblock In {\em Proc. ICDM Workshop}. 344--349.

\bibitem[\protect\citeauthoryear{Zarei, Izadi, and Samani}{Zarei
  et~al\mbox{.}}{2009}]{MinaZarei-NMF:2009}
{\sc Zarei, M.}, {\sc Izadi, D.}, {\sc and} {\sc Samani, K.~A.} 2009.
\newblock Detecting overlapping community structure of networks based on
  vertex-vertex correlations.
\newblock {\em J. Stat. Mech.\/}~{\em 2009,\/}~11, P11013.

\bibitem[\protect\citeauthoryear{Zhang, Wang, and Zhang}{Zhang
  et~al\mbox{.}}{2007a}]{ZhangFuzzy:2007}
{\sc Zhang, S.}, {\sc Wang, R.-S.}, {\sc and} {\sc Zhang, X.-S.} 2007a.
\newblock Identification of overlapping community structure in complex networks
  using fuzzy c-means clustering.
\newblock {\em Physica A\/}~{\em 374}, 483--490.

\bibitem[\protect\citeauthoryear{Zhang, Wang, and Zhang}{Zhang
  et~al\mbox{.}}{2007b}]{NMFZhangShihua:2007}
{\sc Zhang, S.}, {\sc Wang, R.-S.}, {\sc and} {\sc Zhang, X.-S.} 2007b.
\newblock Uncovering fuzzy community structure in complex networks.
\newblock {\em Phys. Rev. E\/}~{\em 76,\/}~4, 046103.

\bibitem[\protect\citeauthoryear{Zhang, Wang, Wang, and Zhou}{Zhang
  et~al\mbox{.}}{2009}]{YuzhouZhang:2009}
{\sc Zhang, Y.}, {\sc Wang, J.}, {\sc Wang, Y.}, {\sc and} {\sc Zhou, L.} 2009.
\newblock Parallel community detection on large networks with propinquity
  dynamics.
\newblock In {\em Proc. SIGKDD Conf.} 997--1006.

\bibitem[\protect\citeauthoryear{Zhao, Zhang, and Pan}{Zhao
  et~al\mbox{.}}{2010}]{CNMFKunZhao:2010}
{\sc Zhao, K.}, {\sc Zhang, S.-W.}, {\sc and} {\sc Pan, Q.} 2010.
\newblock Fuzzy analysis for overlapping community structure of complex
  network.
\newblock In {\em Proc. CCDC}. 3976--3981.

\end{thebibliography}

\begin{received}
Received Month Year;
revised Month Year; accepted Month Year
\end{received}


\end{document}